\documentclass[reprint, amsmath,amssymb,pre]{revtex4-2}

\usepackage{graphicx}
\usepackage{dcolumn}
\usepackage{bm}

\usepackage{xcolor}

\begin{document}

\title{Vertical pullout of a non-spherical intruder from a granular medium: From system-wide response to avalanching around the intruder}

\author{Dominik Krengel}
 \email{dominik.krengel@kaiyodai.ac.jp}
\affiliation{
 Department of Marine Resources and Energy, \\
 Tokyo University of Marine Science and Technology,\\
 4-5-7, Konan, Minato, 108-8477, Tokyo, Japan
}%
\author{Shun Nomura}
\email{nomura.shun@kaiyodai.ac.jp}
\affiliation{
 Department of Marine Resources and Energy, \\
 Tokyo University of Marine Science and Technology,\\
 4-5-7, Konan, Minato, 108-8477, Tokyo, Japan
}%
\author{Shunsuke Ota}
\affiliation{
 Department of Marine Resources and Energy, \\
 Tokyo University of Marine Science and Technology,\\
 4-5-7, Konan, Minato, 108-8477, Tokyo, Japan
}%
\author{Hidenori Takahashi}%
 \email{htak001@kaiyodai.ac.jp}
\affiliation{
 Department of Marine Resources and Energy, \\
 Tokyo University of Marine Science and Technology,\\
 4-5-7, Konan, Minato, 108-8477, Tokyo, Japan
}
\author{Jian Chen}
\email{jchen@jamstec.go.jp}
\affiliation{
 Center for Mathematical Science and Advanced Technology, \\
 Japan Agency for Marine-Earth Sciences and Technology,  \\
 3173-25 Showa-machi, Kanazawa, Yokohama,  236-0001 Kanagawa, Japan
}%

\date{\today}

\begin{abstract}
Intruder mechanics in a granular aggregate is a common subject in engineering and geotechnical applications. However, most studies are limited to spherical intruders or small displacement regimes up to the point of failure. In this work we investigate the vertical pullout of a plate-like intruder buried within a granular aggregate well past the point of failure. While we find the maximum resistance force to depend on the material properties, in the post failure regime the resistance force converges onto the same curve for all friction coefficients. Likewise, the effective geometry of the intruder will always develop the same conical shape on top of the plate, independent of the magnitude of friction, that remains unchanged during the pullout process once established. Further, between the intruder and the aggregate a natural hopper flow develops in which material is transported into the void below the intruder by discrete flow events. 
\end{abstract}

\maketitle


\section{Introduction}
Intruder movement in granular media is a rather common issue in practical considerations, from mixing of materials in e.g. blenders\,\cite{Radl2012,Lehuen2020} to roots\,\cite{Schwarz2010,Yang2021,Milatz2025} and anchoring systems stabilizing soil or structures\,\cite{Liu2012,Sivakumar2013,Liang2021, Kurniadi2025}. The common point is: An foreign object (i.e. the intruder), buried within a granular aggregate, is subject to loading in a particular direction. The surrounding granular matrix initially resists the intruder, up to a limiting force where failure sets in, and the object begins to move through the granular medium. Instead of a loading force some tests also impose a constant velocity on the intruder, but the overall behaviour remains the same. Depending on the problem in question, failure and movement of the intruder is either intended (i.e. for mixing blades) or undesirable (as for soil anchors). Due to the importance for industrial or geotechnical problems, the processes leading to failure and their immediate consequences have been the subject of several studies\,\cite{Trautmann1985,Yu2011,Shah2020,Hossain2020,Jalali2021,Vo2023,Vo2023a}, while movement of an intruder through a granular flow (i.e. the post-failure stage) have been examined in an attempt to formulate a sort of fluid dynamics for granular materials\,\cite{Albert1999,Albert2001,Henann2014,Seguin2016,Takada2019,He2025}.

The vertical pullout of an embedded intruder has been the subject of active research particularly in geotechnics for several decades but is still far from being concluded due to a wide range of complexities involved.
  Based on laboratory experiments, Murray and Geddes\,\cite{Murray1987} and Tagaya et al.\,\cite{Tagaya1988} derived analytical and estimation formulations to predict the peak pullout resistance of an intruder in granular soil. They noted a strong dependence of the pullout resistance on the embedding depth of the intruder, expressed as ratio of \textsl{depth} $H$ to \textsl{intruder diameter} $B$, and the local packing density of the soil. Their analysis generally assumed vertical failure surfaces\,\cite{Trautmann1985} from the boundaries of the intruder to the surface of the aggregate. Newer, more sophisticated experiments instead showed outward facing, curved failure surfaces\,\cite{White2008}, while numerical simulations also showed conical failure surfaces for deep embedded intruders\,\cite{Liang2021}. Liu et al.\,\cite{Liu2012} further showed a dependence of the failure surface on a combination of the embedding depth and local packing density, which was later corroborated by Evans and Zhang\,\cite{Evans2019} and Kurniadi et al.\,\cite{Kurniadi2025}. More recently, Roy et al.\,\cite{Roy2021} extended the existing block-failure based analytical formulations by incorporating a more accurate treatment of the bulk friction angle. Interestingly, Jalali et al.\,\cite{Jalali2021} found a second failure peak during pullout of a spherical intruder from a narrow tube, which cannot be explained in the framework of current models.
  
However, many approaches are limited in their generality: Experimental studies are usually limited by the opacity of the granular medium and have no proper access to the micro-mechanical changes in the aggregate. Analytical solutions can only predict the peak pullout resistance, but do not offer a load-displacement relation for the intruder or the progress of the failure. While continuum simulations give an overview for the evolution of the mean fields of stresses and strains during the pullout, they easily suffer from numerical issues such as mesh-deformation or -rotation, resulting in a loss of convergence and accuracy\,\cite{Nazem2005, Yu2008, Liang2021}. Though it is possible to alleviate these issues with e.g. remeshing techniques, this often incurs significant computational cost, while less rigorous corrections may increase the problems. As the principal objective of pullout simulations is to determine the ultimate pullout resistance, many simulations are therefore only run for small displacement regimes, omitting the post-failure stages of the pullout.

Discrete Element Methods (DEM) allow to investigate the intruder pullout on the particle scale without assuming a priori constitutive relations. Using a combined DEM and Material-Point-Method Liang et al.\,\cite{Liang2021}, showed the formation of rotational bands along the failure surface on the edges of the anchor with differing principal rotation directions below and above the intruder that further weakened the bulk resistance to the intruder motion. Evans and Zhang\,\cite{Evans2019} showed that the force network were mostly independent of the embedding depth. They further showed how the packing density of the aggregate affects the nature of the initial material failure and concluded that in all likelihood the failure cannot be captured by the same equation for all types of granular packings. Vo and Nguyen\,\cite{Vo2023,Vo2023a} linked the pullout resistance of the intruder to the distribution of strong forces acting directly atop of the intruder. Lehuen et at.\,\cite{Lehuen2020} showed that these strong forces are confined to a co-moving conical region above the intruder, regardless of the actual failure surface and decrease almost linearly with decreasing distance from the surface of the aggregate. Lastly, Shah et al.\,\cite{Shah2020} demonstrated that the initial failure during pullout occurs after a series of intermittent changes in the force network around the intruder.

Beyond the initial material failure, multiple studies have focused on the general motion of an intruder through a granular medium. Zhang and Behringer\,\cite{Zhang2017} showed that fluctuations in the force acting on the intruder are the result of simultaneous breaking force chains. Kozlowski et al.\,\cite{Kozlowski2021} found that for a pentagonal intruder stresses are transmitted in a dense cluster in front of the intruder, similar to the observations made in\,\cite{Lehuen2020}. Basak et al.\,\cite{Basak2023} then showed that, for pentagonal intruders more so than for circular intruders, force oscillations also correlate with stick-slip incidents if the intruder is moving through a constant-velocity granular flow. Likewise, Seguin et al.\,\cite{Seguin2016} showed that the granular flow is strongly localized around the intruder, necessitating non-local rheologies\,\cite{Henann2014} to properly describe the flow.

Despite decades of studies across multiple disciplines, understanding the movement of an intruder in a granular medium is still lacking. Results usually are obtained for selected phases of the motion which may be in disagreement even for the same conditions. In this work, we study the motion of a plate-like intruder buried within an aggregate, from the onset of the motion to well beyond the point of failure, where a steady-state like behaviour has been achieved. We show that the failure surface in the aggregate evolves throughout movement of the intruder, and that the shape of this failure surface is linked directly to the resistance of the aggregate against the movement of the intruder. We further show how the flow beyond failure point is correlated with avalanching and jamming in the dilation zone.

\section{Laboratory experiment}
\label{sec:experiment}
\begin{figure}[b!]
 \centering
 \includegraphics[width=0.49\columnwidth]{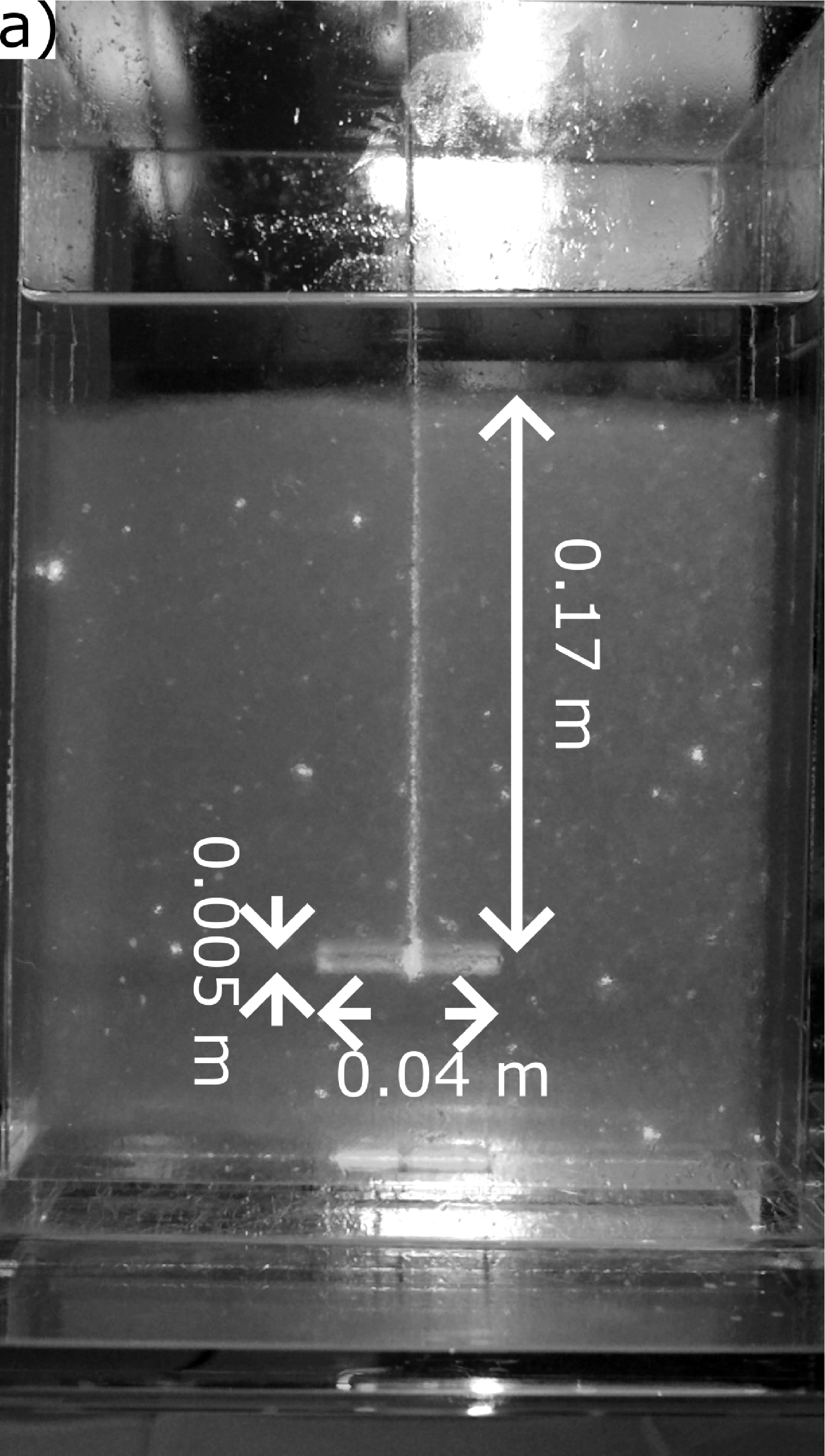}
  \hfill
 \includegraphics[width=0.49\columnwidth]{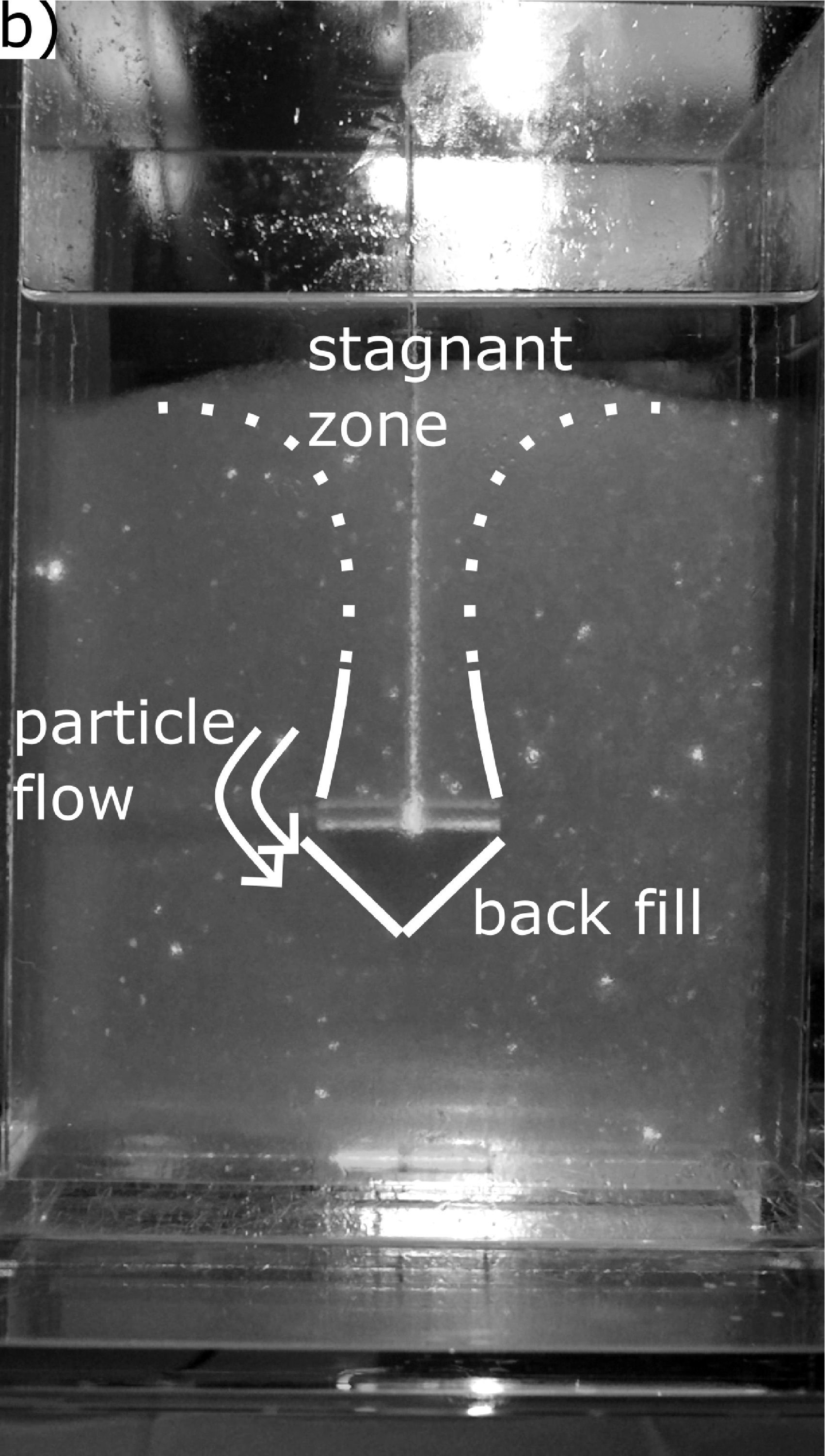}
 \caption{Pictures of the experiment a) before pullout, b) during pullout. Here, $L=0.04$\,m is the width of the plate, $D=0.005$\,m its thickness, and $H=0.17$\,m its initial embedding depth.}
 \label{fig:experiment_pullout}
\end{figure}
To gain insight into the mechanics of the intruder pullout, we perform a simple pullout experiment using a Quartz-Silicone oil mixture. This allows us to directly observe the motion of the intruder inside the aggregate through index matching\,\cite{Ezzein2011, Iskander2015} between the Quartz grains (angular, sharp edged particles; mean radius $\bar{r}=4\cdot 10^{-4}$\,m) and the Silicone oil. In this test we utilize an open rectangular acrylic container with dimensions $0.16\times 0.16\times 0.24$\,m.  For the intruder we use a rectangular resin plate of size $0.04\times0.04\times0.005$\,m. Initially the container is partially filled with particles to a height of  $0.115$\,m,  then the intruder is placed horizontally in the centre of the container and the box filled with particles to the top at $0.24$\,m, see Fig.\,\ref{fig:experiment_pullout},a), resulting in an embedding depth of $h=0.12$\,m. After all particles have been deposited in the box, the plate is being pulled with a constant vertical velocity $v_{\mathrm{plate}} = 10^{-3}$\,m/s by two small metal rods attached to two of its side.  We record the force on the plate with a load cell and record the particle motion by means of a digital camera. 
Figure\,\ref{fig:experiment_pullout} shows the plate before and during pullout. During pullout we observe a stagnant zone on top of the plate that changes shape and size during the pullout process and a stagnant zone below the plate and along the container walls that mostly remains unaffected by the plate movement. Beside the plate particles are continuously transported down into the space opened by the plate uplift. We further observe that the uplifted area does not always have a straight boundary (i.e. failure surface), but rather first bends inwards and then curves outwards as it approaches the free surface (Fig.\,\ref{fig:experiment_pullout}, b) once the plate has moved some distance.

To further assess the influence of material and embedding depth we perform four additional experiments using Toyoura sand (angular, round edged particles; $\bar{r}=9\cdot 10^{-5}$\,m) and Quartz grains under dry conditions. As the influence of the lubrication primarily affects the peak response of the aggregate, and index matching is not possible with Toyoura sand, we thus chose for simplicity's sake to perform these tests in dry conditions. For both materials, the anchor is placed either at an embedding depth of $H=0.12$\,m or at $H=0.17$\,m.

\begin{table}[t]
 \caption{Parameters of the Toyoura sand and the Quartz grains.}
 \label{tab:experiment}
\begin{tabular}{l|l|l}
 Parameter & Quartz & Toyoura sand\\ \hline
 Mean radius $\bar{r}$ [m] & $4\cdot 10^{-4}$ & $9\cdot 10^{-5}$ \\
 Dry density $\gamma^{\prime}$ [$10^{3}$ N/m$^3$] & $15.42$ & 15.45\\
 Internal friction angle $\Phi$ [$\circ$] & 42 & 40.2
\end{tabular}
\end{table}

\begin{figure}[b]
 \centering
 \includegraphics[width=\columnwidth]{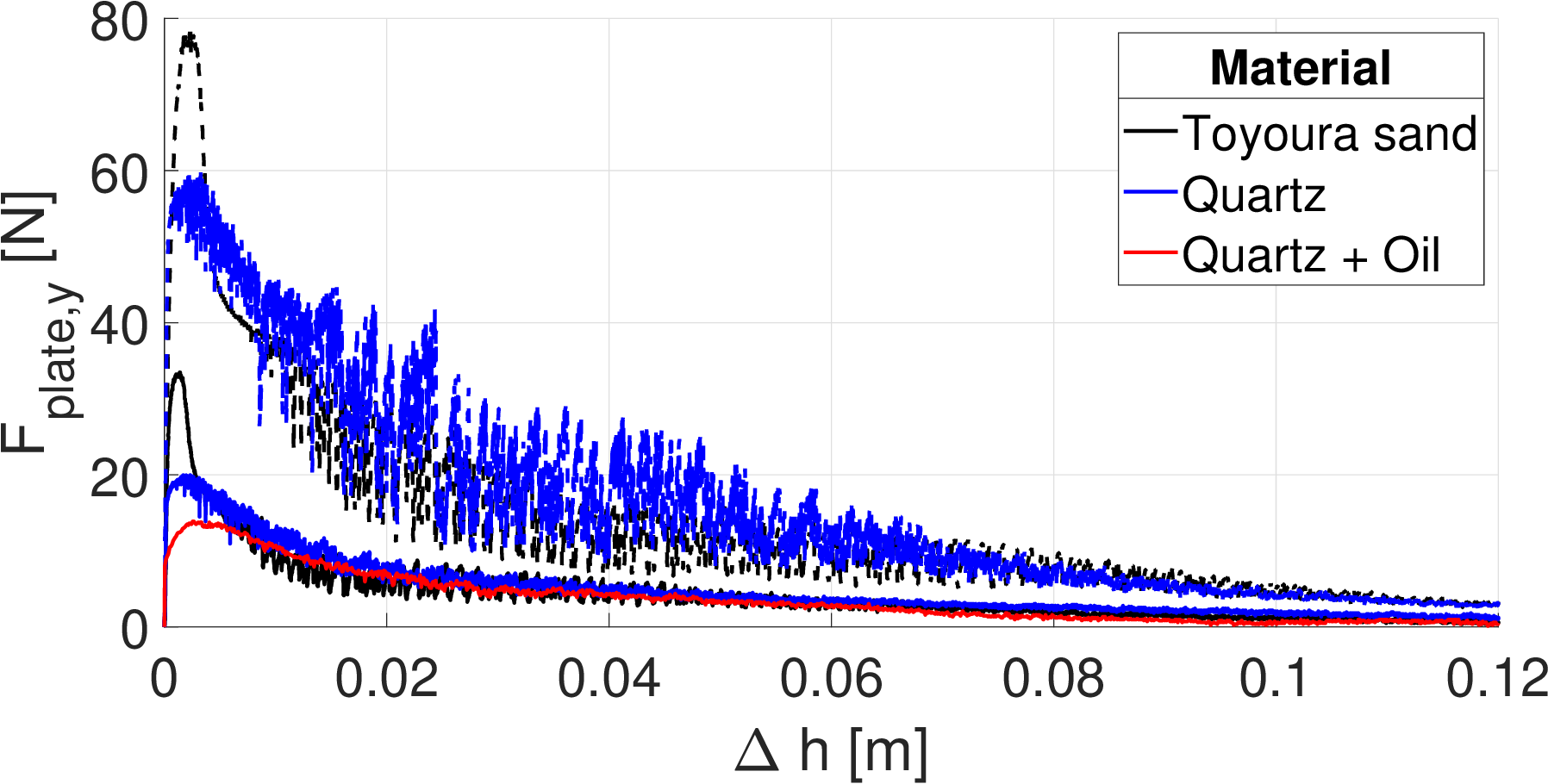}
\caption{Experimental curve for the pullout resistance $F_{\mathrm{plate},y}$ vs. displacement $\Delta h$ for the plate pullout with different granular materials. The solid lines show data for an embedding depth $h = 0.12$\,m, the dashed lines for $h= 0.17$\,m. Quartz-Oil mixture data exists only for $h=0.12$\,m.}
 \label{fig:experiment_Force}
\end{figure}

Figure.\,\ref{fig:experiment_Force} shows the resistance-displacement curve for different materials. The overall behaviour is typical for a stress-strain relation in granular materials, with a non-constant confining pressure in the `steady-state' at the end. We can broadly distinguish three different regimes: A quick shear hardening at the beginning, followed by an equally fast shear softening until around $\Delta h=0.01$\,m. Secondly, a noisy transition regime in which the resistance continues to decay in a non-linear manner until around $\Delta h=0.06$\,m for an embedding depth of 0.17\,m. Finally, a `steady state' is reached, where the resistance is decreasing almost linear to the equally linear decrease in the overburden (confining) pressure with increasing pullout distance. The transition between the three different regimes appears to occur earlier if the initial embedding is shallower.
Given the same depth, the pullout resistance for dry quartz grains shows a lower maximum than for Toyoura sand, but appears to be converging against almost the same `steady state' limit. Increasing the embedding depth leads to a higher peak resistance as well as higher noise. Lubricating the quartz grains with silicone oil instead decreases the pullout resistance to around a quarter of its unlubricated strength and brings the steady state values closer to those of shallower embedding depths. In the `steady-state', the pullout resistance appears to converge against similar values regardless of initial embedding depth or material, but our current experimental data is insufficient to give a definite answer.

Various analytical formulations exist to predict the peak pullout resistance (e.g. Murray and Geddes\,\cite{Murray1987} and the monograph by Das and Shukla\,\cite{Das2013}), based on different a-priori assumptions on the shape of the failure surfaces. This difference between the assumed and the actual shape of the failure surface then is a common source of deviation between the analytical and measured pullout resistance, see also e.g.\,\cite{White2008,Roy2021}. In addition, while the analytical apparatus can give an estimate of the maximum pullout resistance it does neither provide insight into the evolution of $F_{\mathrm{plate},y}$ nor the physical reason for the different regimes (hardening and softening, noisy transition, `steady-state'). As the particle-scale is difficult to access by experimental approaches, we instead use a discrete element simulation to study the micro-mechanics of the plate pullout.

\section{DEM Simulation}
\label{sec:simulation}
To investigate in detail the micro-mechanical response of the aggregate to the intruder pullout, we conduct several Discrete Element Method (DEM) simulations. We focus primarily on the immediate response of the material after the onset of the pullout until a stable state has been reached, not on the complete pullout of the intruder from the granular medium, nor on the first failure of the aggregate. While the experiment and theory describe the pullout process in a three-dimensional medium, we instead conduct two-dimensional simulations, where the system can be considered as a vertical slice through the real setup. This allows significantly faster computation than in three dimensions for equivalent system sizes, in particular for non-spherical particles. In addition, the experiment with the Quartz-Silicone oil mixture provided a quasi-2D visualisation of the grain dynamics around the intruder, which provides a suitable reference point in interpreting the particle dynamics in our simulation.

\subsection{Methodology}
In this study we use a two-dimensional `hard-particle soft-contact' DEM based on the monograph by Matuttis and Chen\,\cite{Matuttis2014}. Particles are represented by convex polygons instead of discs, as real particles are generally aspherical and the particle shape is known to significantly affect the stress response of granular aggregates\,\cite{Athanassiadis2014,Krengel2023}. 

For two particles $A$ and $B$ in contact, the tangential direction $\mathbf{t}$ is given by the intersection line between the two polygons, which also fixes the normal direction $\mathbf{n}$. The elastic force
\begin{equation}
 F_{\mathrm{el}}=\frac{Y\cdot A}{l}
\end{equation}
is proportional to the overlap area $A$ and the Young's modulus $Y$. A characteristic length
\begin{equation}
 l=4\frac{|\mathbf{r}_{A}|\cdot|\mathbf{r}_{B}|}{|\mathbf{r}_{A}|+|\mathbf{r}_{B}|},
\end{equation}
based on the contact vectors $\mathbf{r}_{A,B}$ between the centres of mass and the centroid of the overlap area, is required to define the normal force in units of [N]. Further, a dissipative force
\begin{equation}
 F_{\mathrm{diss}}=\gamma \sqrt{mY}\frac{\dot{A}}{l},
\end{equation}
with the reduced mass $1/m = 1/m_{A} + 1/m_{B}$ and a damping constant $\gamma$, is acting in normal direction, so that the total normal force is
\begin{equation}
 F_{N} = F_{\mathrm{el}}+F_{\mathrm{diss}}.
\end{equation}
A friction force in tangential direction is implemented through the Cundall-Strack model\,\cite{Cundall1979},
\begin{equation}
 F_{\mathrm{T}}(t) = \left\{\begin{array}{ll}
        F_{\mathrm{T}}(t-\mathrm{d}t)-k_{\mathrm{t}}v_{\mathrm{t}}\mathrm{d}t & |F_{\mathrm{T}}(t)|\leq \mu F_{\mathrm{N}}\\
				\mathrm{sgn}\left(F_{\mathrm{T}}(t-\mathrm{d}t)\right)\mu F_{\mathrm{N}} & |F_{\mathrm{T}}(t)| > \mu F_{\mathrm{N}}
\end{array}\right.,
\end{equation}
where $v_{\mathrm{t}}$ is the relative tangential velocity at the contact point and $k_{\mathrm{t}}$ the `tangential stiffness' of a spring. The total force $\mathbf{F} = F_{\mathrm{N}}\mathbf{n} + F_{\mathrm{T}}\mathbf{t}$ acts upon the centroid of the overlap area which induces a torque on each particle as
\begin{equation}
 \bm{\tau}_{\mathrm{A,B}} = \mathbf{r}_{\mathrm{A,B}}\times\mathbf{F}.
\end{equation}

\subsection{Setup}
\begin{figure}[b!]
 \centering
 \includegraphics[width=\columnwidth]{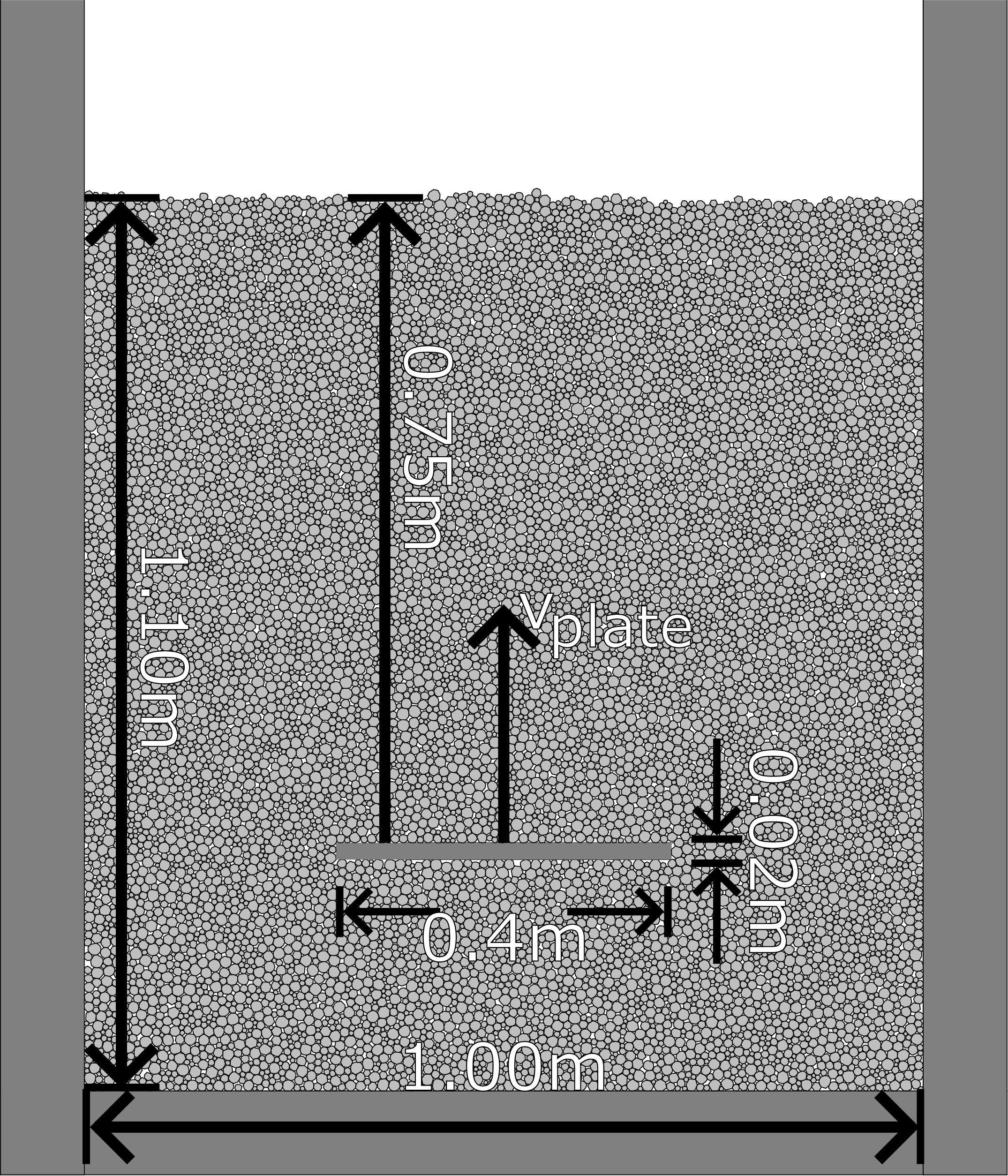}
\caption{Initial configuration of the pullout test. The lateral walls and the bottom plate remain fixed while the plate is slowly pulled out with constant velocity  $v_{\mathrm{plate}}$.}
 \label{fig:anchor_setup}
\end{figure}

The particles are modelled as irregular, convex polygons with 6 to 13 corners and a uniform circum-radius distribution with mean radius $\bar{r}=0.005$ m and a radius variation of up to 50\%. Compared to the experiment, the particle size is chosen one order of magnitude larger to allow feasible simulation times. The plate which acts as the intruder is modelled as a rectangular polygon with sidelength $0.4$\,m by $0.05$\,m or approximately 40 by 2 particle diameters. We generate the initial deposit by first dropping 3000 particles into an empty box under gravity. The configuration is smoothed and the plate placed is centred on top of the particle layer. Then about 9000 more particles are deposited above the plate so that the plate is buried at $26.7\%$ of the assembly height, $0.75$\,m below the surface, see Fig.\,\ref{fig:anchor_setup}. 

Compared to the experiment, the intruder in the simulation is placed much deeper within the aggregate. As the particle size is much larger, we intended to have a sufficient amount of particles above the intruder than can be displaced during the pullout process.
During deposition the particles are frictionless to generate a dense initial packing, while the actual pullout is performed with finite friction. After deposition, the simulation of the pullout starts where the plate is slowly lifted upwards against gravity for 20 seconds with $v_{\mathrm{plate}}=0.01$\,m/s, i.e. half the plate thickness per second . During pullout the plate has no material response and only moves upwards with constant velocity without horizontal movement or rotation. Further simulation parameters are given in tab.\,\ref{tab:materialparameters}. As the simulations are performed in two dimensions, the dimensions of the material density $\varrho$ and the Young modulus $Y$ are given in units of [kg/m$^2$] and [N/m] to allow definition of the force in units of [N]. To confirm reproducibility, we perform two additional simulations for $\mu=0.5$ and $v_{\mathrm{plate}}=0.01$\,m/s, which we discuss in more detail in the appendix.

\begin{table}[h]
 \caption{Numerical parameters for the simulations.}
 \begin{tabular}{lll}
Parameter & & Value\\
\hline
 Num. particles  & $N$   & $\sim 12000$\\
 Coeff. friction & $\mu$ &  $[0.1,0.25,0.5,1.0]$\\
 Young modulus   & $Y$   & $10^{9}$ [N/m]\\
 Damping parameter & $\gamma$ & $0.5$\\
 Material density & $\varrho$ & $5000$ [kg/m$^2$]\\
 Mean particle radius & $\bar{r}$ & $0.005$ [m]\\
 Particle diameter range & & $[0.5\ 1.5] \bar{r}$\\
 Plate size      & $l_{\mathrm{plate}}\times d_{\mathrm{plate}}$ & $0.4\times0.02$ [m]\\
                 &                                       & $\widehat{=}80\times4$ [$\bar{r}$]\\
 Pullout velocity & $v_{\mathrm{plate}}$ & $0.01$ [m/s]\\
                  &                      & $\widehat{=} 2$ [$\bar{r}$/s]\\
 Timestep        & $\mathrm{d}t$ & $2.5e-6$ [s]\\
 Simulation time & $t_{\mathrm{max}}$ & 20 [s]\\
\hline
 \end{tabular}
 \label{tab:materialparameters}
\end{table}

\section{Results}
\label{sec:results}
\subsection{Pullout resistance of the plate}
\begin{figure}[t]
 \centering
 \includegraphics[width=\columnwidth]{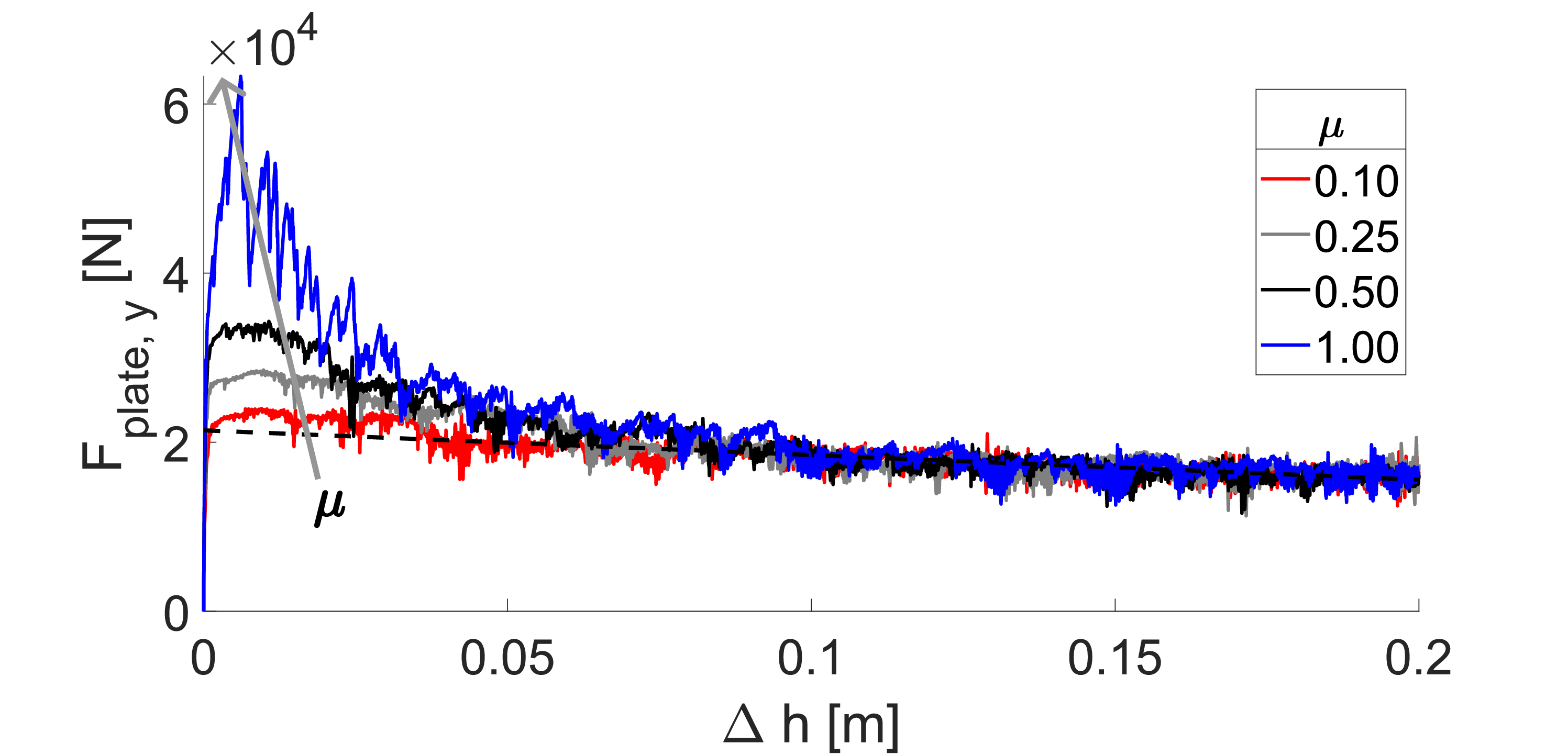}
\caption{Force-displacement diagram of the plate with different coefficients of friction $\mu$. The dashed lines mark the linear fits in the range $\Delta h =[0.08\ 0.2]$. The grey arrow indicates the variation of the occurrence and magnitude of peak resistance with $\mu$.}
 \label{fig:anchor_force_mu}
\end{figure}

When the intruder is pulled out of the aggregate, it experiences a resistance-force (i.e. a drag-force) against its movement, see Fig.\,\ref{fig:anchor_force_mu}. Compared to the experiment, the resistance force in the simulation is two orders of magnitude larger due to the deeper embedding depth in the simulation. In the simulation we observe the same three regimes of the pullout resistance as in the experiment: An initial increase (shear hardening) as the material compacts is followed by a strong but short drop in resistance (shear softening) when the granular aggregate dilates (phase 1). With increasing pullout distance $\Delta h$, the rate of the stress drop decreases non-linearly (phase 2). At the end (phase 3), the stress-displacement curve decreases linearly. The exact location of the transition between the different phases depends on the actual material properties of the aggregate, such as the friction coefficient $\mu$ or particle shape. Overall, the evolution of the pullout force with pullout distance follows the typical stress-strain response of granular aggregates (e.g.\,\cite{Krengel2023, Krengel2025}), with the continuous decay in the `steady-state' in phase 3 being due to the absence of a constant confining pressure on the system boundaries. Further, compared to usual bi- and triaxial tests, the loading originates from inside the aggregate, and not from the boundaries.

As shown in Fig.\,\ref{fig:sample_selection}, we can separate the different phases approximately by the curvature $F''$ of $F_{\mathrm{plate}}(\Delta h)$: In phase 1 $F_{\mathrm{plate}}(\Delta h)$ is concave, i.e. $F'' < 0$. In Phase 2  $F_{\mathrm{plate}}(\Delta h)$ is convex and $F'' > 0$. Lastly, in phase 3, the functional form of $F_{\mathrm{plate}}(\Delta h)$ is linear, and the curvature vanishes, $F'' = 0$. The transition between phases 1 and 2 occurs later for lower values of $\mu$ while the transition between phases 2 and 3 appears to occur slightly earlier. 
We omit computing the exact transition step between different phases as the data presented here is too noisy and derivation would only amplify the noise. Instead we provide a rough estimate of the transition between phases 1 and 2 based on visual analysis of the curves in Tab.\,\ref{tab:transition_1to2}. 
As the transition between phases 2 and 3 appears to be almost independent of $\mu$ for simplicity we set the transition at $\Delta h = 0.08$\,[m] in the rest of this work  when all curves have entered the linear phase. We will provide a deeper analysis of the different phases and their dependencies in a follow up work.

\begin{table}[h]
\centering
\caption{Manual estimation of the displacement for the transition between phases 1 and 2.}
\begin{tabular}{l|c|c|c|c}
 $\mu$ & 0.10 & 0.25 & 0.50 & 1.00\\\hline
 $\Delta h$\,[m] & 0.035 & 0.025 & 0.020 & 0.015
\end{tabular}
\label{tab:transition_1to2}
\end{table}

As with previous biaxial compression tests\, \cite{Krengel2025}, we observe that the peak pull out resistance scales with the friction coefficient $\mu$, i.e. depends on the properties of the granular material used. In contrast, in phase 3 the stress-displacement curve collapses onto the same form for all $\mu$, thus has some independence of the material properties, and can be approximated with a simple linear fit as
\begin{equation}
 F_{\mathrm{linear}} = (-2.92\pm0.51)\cdot10^{4}\cdot \Delta h + (2.14\pm0.09)\cdot10^{4}.
 \label{eq:Fresist_dh}
\end{equation}

Similar to our results, Shah et al.\,\cite{Shah2020} also found a dependence on $\mu$ for the peak pullout resistance of a deeply embedded intruder. Additionally, we observe that the peak pullout resistance is reached later if $\mu$ is lower, with the trend being indicated by the arrow in Fig.\,\ref{fig:anchor_force_mu}.

Results from earlier biaxial compression tests with monodisperse shape distributions suggest that the linear regime of the force-displacement curve may also stratify for increasing $\mu$ for pullout tests with monodisperse particle shapes (see e.g.\,\cite{Binaree2020,Krengel2025}). On the other hand, as we are in the frictional drag regime\,\cite{Rognon2025}, the stress response is rate independent for $v_{\mathrm{plate}}=[0.001,0.02]$ m/s, and the deformation is quasi-static in this velocity range (see also e.g.\,\cite{Albert1999}).

\subsection{Macroscale response}
\begin{figure}[b!]
 \centering
 \includegraphics[width=\columnwidth]{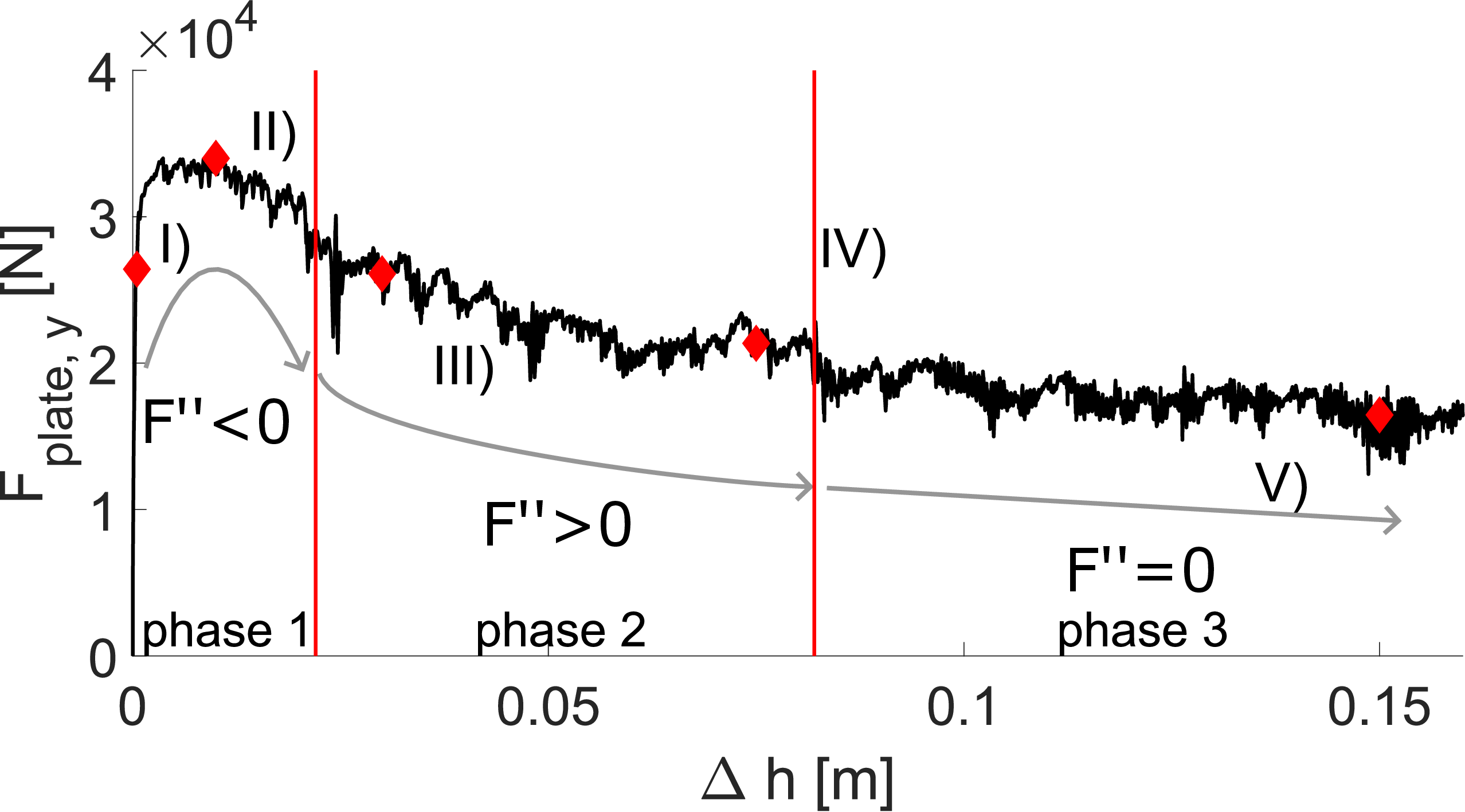}
 \caption{Selection for the timesteps I)-V) for the macroscopic fields at different stages of the pullout ($\mu=0.5$). The grey arrows mark the curvature of $F_{\mathrm{plate}}(\Delta h)$ which is used to separate the pullout into different phases.}
 \label{fig:sample_selection}
\end{figure}

In order to evaluate the bulk properties of the aggregate, we discuss five different steps during the pullout, I)--V), see Fig.\,\ref{fig:sample_selection}. Step I) was chosen at $\Delta h_{I} = 5.0\cdot10^{-4}$\,[m] during initial shear hardening, shortly after the onset of plate movement. Step II) is located during peak pullout resistance at $\Delta h_{II} = 1.0\cdot10^{-2}$\,[m]. Steps III) and IV) are near the beginning and end of phase 2, at $\Delta h_{III} = 3.0\cdot10^{-2}$\,[m] and $\Delta h_{IV} = 7.5\cdot10^{-2}$\,[m]. Lastly, we selected step V) during the `steady-state' in phase 3 at $\Delta h_{V} = 1.5\cdot10^{-1}$\,[m]. In total, we have two representative steps from phase 1, two steps from phase 2 and one step from phase 3. Although the transition between phases is influenced by the friction coefficient $\mu$, for consistency with regard to the pullout distance $\Delta h$, we keep the same selection of steps for tests with different $\mu$.

For each of these steps we determine the porosity $n$ field, the vertical velocity field $v_{y}$ and the force chains $\propto F_{N}$, and the cumulative particle orientation $\theta$ field, which we will discuss in detail below.

\begin{figure*}[t]
\centering
 \includegraphics[width=\textwidth]{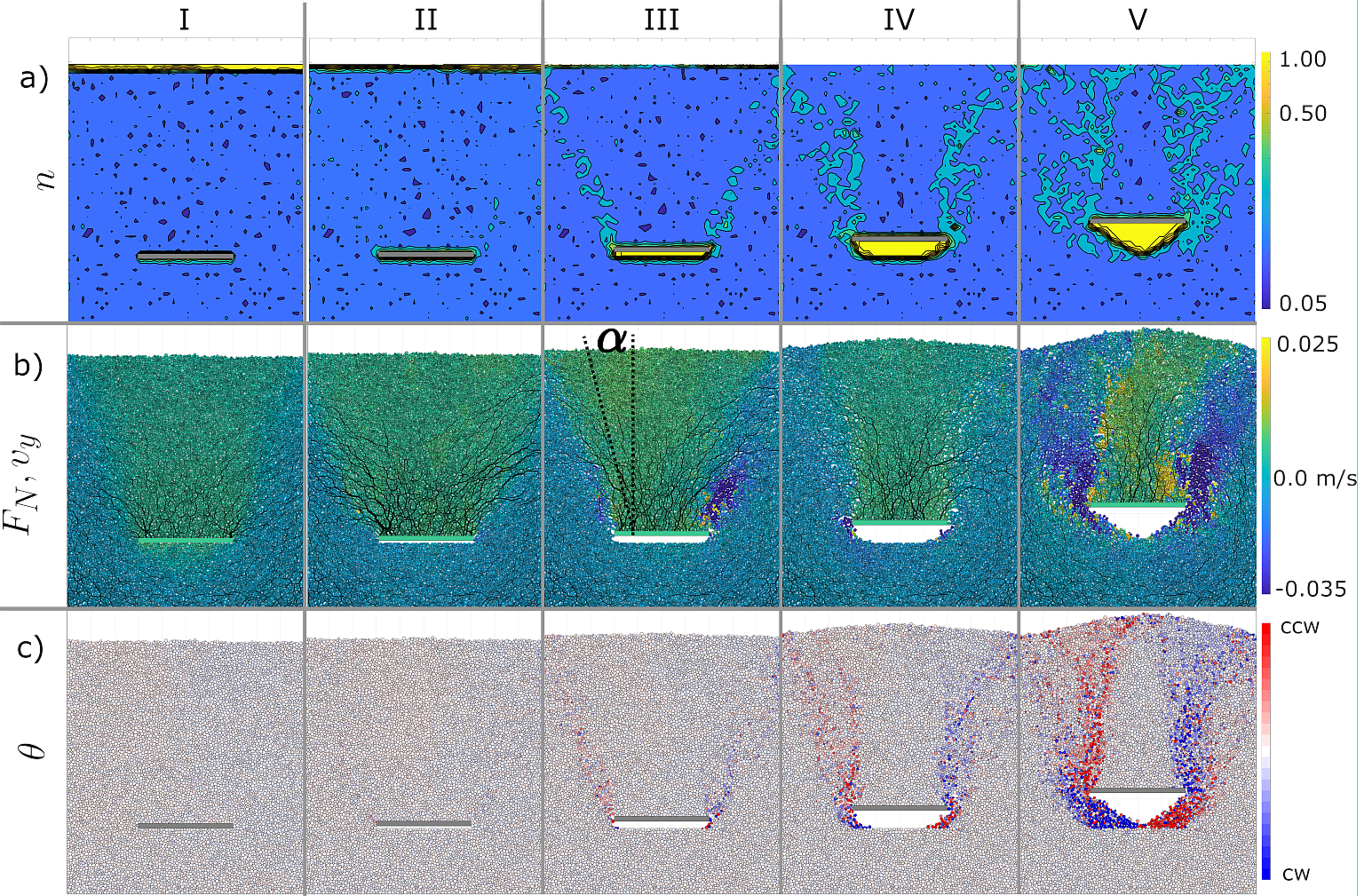}
 \caption{System-wide responses at various stages of the plate-pullout ($\mu=0.5$): a) Local porosity $n$, where blue colouring indicates low porosity and yellow indicates high porosity. b) Vertical velocity field $v_{y}$ with superimposed force chains $\propto F_{N}$. Green particles move upwards at the same speed as the plate, yellow particles move faster, and blue particles move downwards relative to the plate. c) Cumulative particle orientation $\theta$. Clockwise rotation (CW) is marked in blue, counter-clockwise rotation (CCW) in red. $\alpha$ is the approximate inclination of the failure surface with respect to the vertical direction.}
 \label{fig:bulk_fields_small}
\end{figure*}

\subsubsection{Porosity}
Figure\,\ref{fig:bulk_fields_small} a) shows the bulk porosity
\begin{equation}
 n=1-\frac{\sum{A_{\mathrm{particles}}}}{A_{\mathrm{box}}}
 \label{eq:porosity}
\end{equation}
 in the aggregate for the different sampling steps. Initially (step I) the porosity is homogeneous throughout the aggregate. Due to the lack of confining pressure $n$ decreases at the free surface on top of the aggregate. As the aggregate is initially very dense, the onset of the intruder movement (step II) leads to dilation rather than compaction, visible as an increase in porosity in the immediate vicinity of the plate. During failure, in phase 2, the material begins to dilate along diagonal failure surfaces pointing up- and outwards from the plate (step III). The upward movement of the plate leaves an empty space below the intruder. Particles to the left and right of the plate flow into this space, which then increases the porosity left and right of the plate. With increasing pullout distance (step IV), the diagonal dilation zones left and right of the intruder increase in size, while the dense zone on top of the plate shrinks. 
Finally, in the `steady state' (step V), the dilation zone has expanded throughout the aggregate into a bowl-like shape. Only the aggregate below and to the side of the initial plate location, and a conical region on top of the plate remain unaffected. Overall, the boundaries of the stagnant zones on each side of the intruder tend to resemble a hopper with the typical diagonal walls and a narrow gap at the bottom.
The porosity of the aggregate also correlates with particle translation and rotation: In a dense medium translation and rotation are restricted due to confinement by other particles, but in a porous medium individual particles can move more freely.

\subsubsection{Velocity field and force chains}
\label{sec:sub_velofields}
As the plate moves upwards, it pushed the particles in its path, which in turn push higher particles along their own contact network. In Fig.\,\ref{fig:bulk_fields_small} b), we show the vertical particle velocity $v_{y}$ with superimposed force chains ($\propto F_{N}$). Initially, during shear hardening (step I), the plate uplifts a trapezoidal wedge of particles with straight edges, see also\,\cite{Roy2021}. Force chains begin to emanate from the plate, with the strongest force acting predominantly on the left and right sides on top of the plate and spreading diagonally outwards along the surface of the uplift area. Below the plate, a small section of the aggregate also moves upwards due to the reduction in their local overhead pressure, leading to a localized relaxation of the aggregate. 

Until maximum pullout resistance (step II), the uplifted particle wedge increases in size. The force chains extending from the plate increase in strength and length. The strongest forces still act on the edge of the plate and point outwards, but now the force network in-between the edges also begins to increase in strength. The free surface on top of the aggregate is still flat, no heap has been formed yet; any dilation from the plate movement has been compensated by compaction elsewhere within the aggregate. At peak resistance, the material fails along the boundary of the uplift area, which marks its failure surface. 

After shear failure (step III), the uplifted segment shrinks laterally. The shrinking starts close to the plate and progresses upwards, leading to a curved failure surface. In this stage the force chains start to point upwards instead of outwards and to move closer to the plate centre. In the dilated area along the failure surface discrete, fast downward particle flow begins to occur. The extend and magnitude of the downflow varies with each instance of flow. The flowing particles are deposited below the plate, slowly refilling the empty space.

By step IV), the compact uplift zone has turned rectangular, with some larger area still remaining far from the plate. Internal compaction no longer compensates for the extent of the dilation and a heap begins to form on the free surface on top of the aggregate. The strong force chains have now migrated towards the plate centre and only weak forces remain on the plate edges. The expansion of the dilation zone leads to stronger particle downflow, which slowly erodes the stable parts of the aggregate that had previously been unaffected by the intruder uplift through kinetic particle impacts.

In the steady state, step V), the shape of the uplift zone has stabilized into a conical structure with mostly straight edges and an almost constant slope angle on top of the plate matching the shape of the dense region seen in the porosity $n$. The force chains remain stable in the centre of the plate. Strong avalanching occurs left and right of the stable zone, with the mobilized area reaching until the free surface of the aggregate. Particle downflow may also cause part of the aggregate on the outer and inner failure surface to temporarily move upwards.

As in the experiment, we observe curved boundaries of the uplifted material, in particular after shear failure in phase 2. The extent of this curvature is weaker than in the experiment, possibly due to the differences in particle size. However, the shape of these boundaries changes throughout the failure process and after, and thus it is difficult to assign one specific shape of the failure surface to the material failure itself. In phase 3, in the steady-state, the curved failure boundaries have mostly disappeared due to erosion and avalanching along the cone surface, but may momentarily re-appear depending on the fluidisation of the aggregate  during an avalanche.

\subsubsection{Cumulative particle rotation}
Besides rectilinear motion, the upward movement of the plate also induces rotation in the particles. In Fig.\,\ref{fig:bulk_fields_small} c) we plot the cumulative particle orientation changes $\theta$ with respect to their initial orientation. During shear hardening and peak shear (steps I and II), the aggregate is too dense for the particles to rotate. As the failure surface begins to emerge (step III), particles along the failure plane are increasingly less constrained by the surrounding granular matrix and start to rotate, with the direction of rotation dependent on the direction of the rectilinear motion.
 The magnitude of rotation increases as the dilation zones spread (step IV and V). Particle rotation generally follows along both failure surfaces of the dilation zone between the two stagnant areas. Below the plate, the rotation direction reverses for the deposited particles: particles that rotated counter-clockwise above the plate now show clockwise rotation and vice-versa. Similar organisation into zones of preferential direction along the failure lines was also observed by Liang et al.\,\cite{Liang2021} with MPM simulations. At the end, in the steady state, a counter-rotating zone appears to be forming outside the rotation zone, but our data is insufficient to mark it clearly.

\subsubsection{Influence of inter-particle friction}
\begin{figure}[t]
 \centering
 \includegraphics[width=\columnwidth]{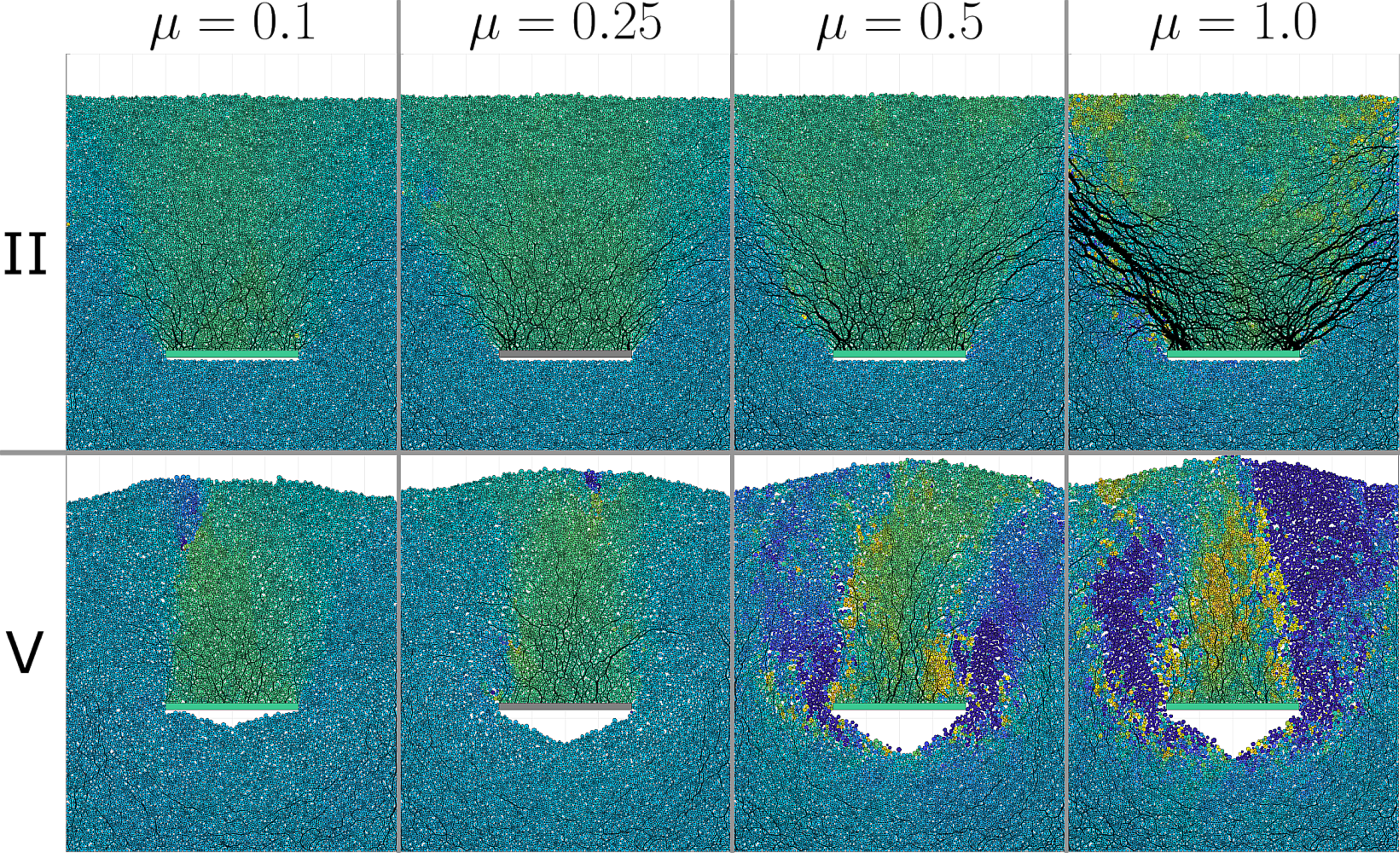}
 \caption{Influence of the friction coefficient $\mu$ on the velocity field $v_{y}$ and force chains $\propto F_{N}$ during pullout step II and pullout step V. Green particles move upwards at the same speed as the plate, yellow particles move faster, and blue particles move downwards relative to the plate. See Fig.\,\ref{fig:bulk_fields_small} b) for the corresponding colourscale.}
 \label{fig:bulk_velocityfield_mu}
\end{figure}

Friction strongly influences the stability of intergranular contacts: In general, for lower values of $\mu$ an aggregate is more susceptible to applied forces and the granular matrix deforms more easily, while conversely, for large $\mu$ the opposite is true. In Fig.\,\ref{fig:bulk_velocityfield_mu} we show the velocity field and force chains for different values of $\mu$ at pullout steps `II' and `V'. Initially, the zone of influence of the plate is much smaller for lower values of $\mu$, and the stagnant zone appears more rectangular (similar to the assumptions of older analytical attempts to predict the peak shear resistance as in e.g.\,\cite{Trautmann1985}), and the force network does not propagate as far as for large values of $\mu$, which is within expectation. As $\mu$ increases the zone of uplifted material increases up to a limit, but the force chains going out from the plate and the particle velocities in the zone continue to increase in strength. In particular, for $\mu=1.0$, the force chains reach the fixed outer walls of the container. In this case the walls provide additional resistance to the pullout, however the uplift wedge is truncated, which reduces the pullout resistance again.
In contrast, In the `steady state' regime, the stagnant zone on top of the plate appears to be the same triangular shape, independent of $\mu$. While the length and strength of the force chain still shows influence of $\mu$ the force chains are always contained within the stagnant area. To summarize, while the maximum shear resistance is dependent on the friction coefficient $\mu$, respective the size of the granular domain, once the granular flow has been established, after shear failure, the influence of $\mu$ and the domain size vanishes.
In addition, we also observe, that both the slope angle of the backfill area below the plate, and the slope angle of the heap forming on top of the aggregate show some relation with $\mu$. In the backfill, the slope becomes steeper with increasing $\mu$, matching the dynamic angle of repose in granular flows. Lastly, we find that particles outside the uplift wedge appear to move faster if $\mu$ is large.

As $\mu$ controls the stability of contacts between the particles, the zone of influence of the plate expands as $\mu$ increases. That means, for larger $\mu$, the failure surfaces develop faster and stronger as inter-particle voids opened by the uplift are less likely to be filled by downward sliding particles. The earlier onset of dilation corresponds to the earlier occurrence of peak resistance in Fig.\,\ref{fig:anchor_force_mu}. A similar effect can also be observed for biaxial compression tests in e.g.\,\cite{Binaree2020,Krengel2025}, but is not discussed in either work. As the pullout progresses weakly frictional particles ($\mu = 0.1$) develop only small dilation zones with little variation in porosity, including below the plate. For low $\mu$ the contacts between particles are not stable enough to prevent particles filling up any gaps in the lower layers, which also explains the smaller heap on top of the aggregate. On the other hand, for materials with larger friction ($\mu\geq0.5$), the maximum extent of the dilated zone is largely the same, forming a hemispherical region below the plate, where the displaced material is deposited, and then extending diagonally outwards above the plate.

\subsection{Dead load: The effective shape of the intruder}
\label{sec:shape_intruder}
Although we model the intruder as a plate, as Fig.\,\ref{fig:bulk_fields_small} has shown, there exists a co-moving stagnant zone on top of the plate between the two failure surfaces. This stagnant zone, sometimes also referred to as dead load, is often treated separately from the intruder in literature. 
However, as there is essentially negligible relative motion between the stagnant zone and the intruder, we can regard it as an extension of the intruder and thus define the whole co-moving region as the effective shape of the intruder.

Initially, in phase 1 and the first half of phase 2 (Fig.\,\ref{fig:bulk_fields_small}, steps I)--III), the plate uplifts a trapezoidal area of the aggregate, with the vertical edges of the zone pointing outwards and upwards from the plate edges. 
As $\Delta h$ increases (steps III)--IV), the stagnant zone becomes rectangular (second half of phase 2) due to the sides of the uplift wedge eroding away under gravity and particle impacts. This rearrangement of the granular matrix starts at the plate and gradually moves upwards, leading to curved failure surfaces.
Finally, in the linear regime in phase 3 (step V), the stagnation zone above the plate stabilizes into a conical shape with near constant slope angles $\alpha$, see Fig.\,\ref{fig:bulk_fields_small} b). 
\begin{figure}[b]
 \centering
 \includegraphics[width=\columnwidth]{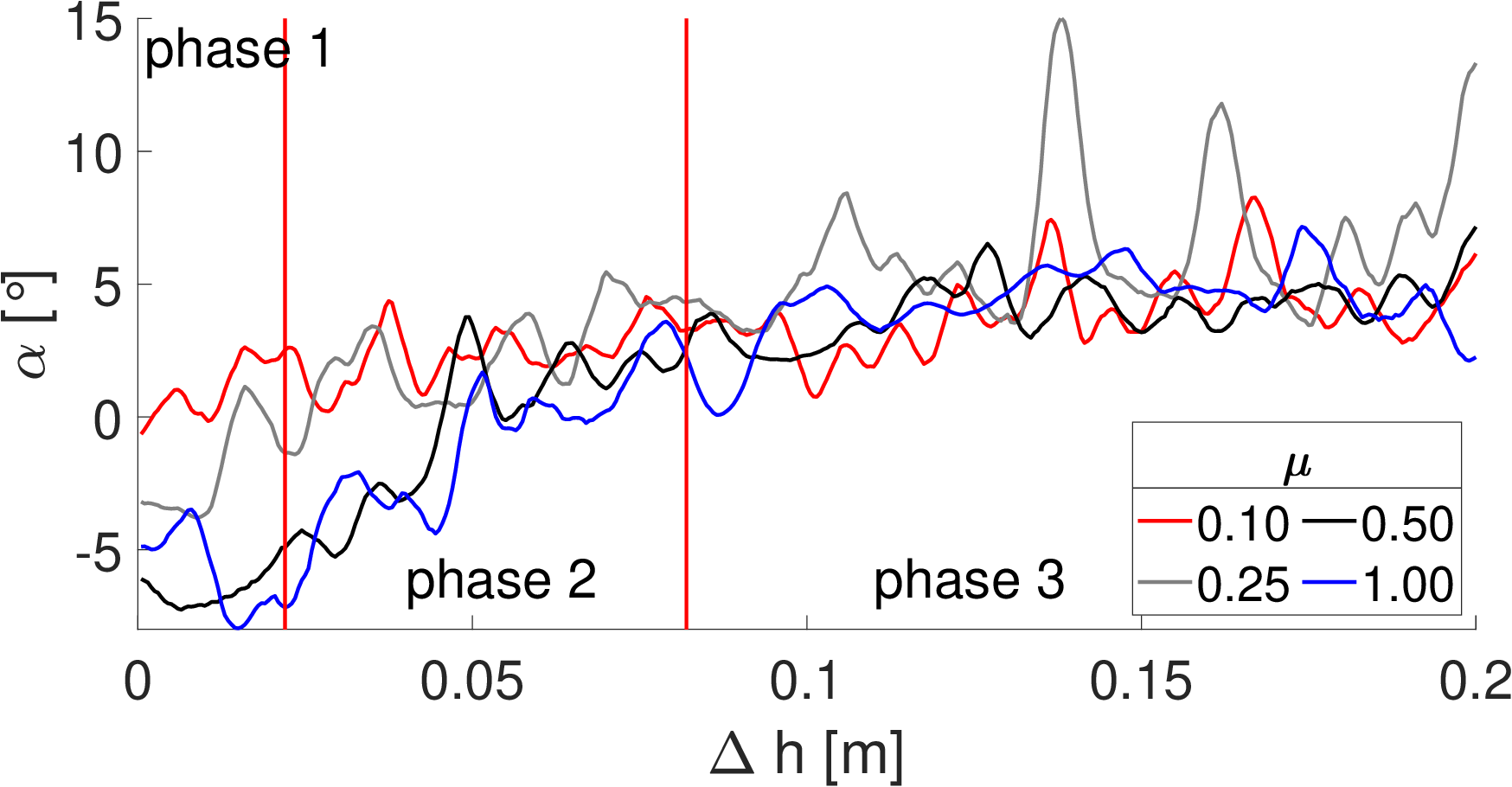}
 \caption{Evolution of the inclination angle $\alpha$ of the sides of the uplifted volume with pullout distance $\Delta h$.}
 \label{fig:anchor_shape}
\end{figure}

To measure the shape of the stagnation zone, we first filter out all particles that lie below the plate, then all particles with $v_y < v_{\mathrm{plate}}$, and then all remaining particles that have no force path connecting them to the plate. For the remaining structure we can then measure the mean horizontal particle coordinate by height above the plate,  to the left and right of the plate centre. The mean horizontal coordinate, together with the upper two vertices of the plate, provides an approximation of the cone side angle $\alpha$, where negative values indicate that the slope points outwards, positive values represent the slope pointing inwards.
This measure does not provide the exact slope angle, as the instantaneous velocity is prone to strong fluctuations, and thus the surface of the resulting structure rather uneven. Filtering by the cumulative particle rotation $\theta$ (Fig.\,\ref{fig:bulk_fields_small} c) can in principle isolate the failure surface much better, but is limited to higher displacements where enough rotation has accumulated and so does not work in early stages of the pullout. However, as we are only interested whether the co-moving structure is pointing outward or pointing inwards with respect to the plate edge, our chosen method is both sufficiently accurate and easily automatable. A more detailed explanation on the computation of $\alpha$ is given in the appendix.

Figure\,\ref{fig:anchor_shape} shows the evolution of $\alpha$ with pullout distance $\Delta h$ for different values of $\mu$. For $\mu=0.5$ and $\mu=1.0$, the co-moving structure is initially oriented outwards, while for $\mu=0.1$, the structure is initially closer to a rectangle with vertical sides, as already observed earlier in Fig.\,\ref{fig:bulk_velocityfield_mu}.
 At the end of phase 2 ($\Delta h \geq 0.05$\,[m]) $\alpha$ converges onto the same value for all $\mu$, i.e. towards the same shape,
\begin{equation}
 \alpha_{\mathrm{stable}} = 4.11\pm0.14.
\end{equation}
 We understand this as the effective shape of the intruder after failure not being determined by frictional interactions, but through other means. Due to the independence from $\mu$, the slope angle in the stable regime also differs from the angle of repose for free standing granular heaps. In contrast to the slopes of the uplift zone, the slopes of the backfill below the plate, shown in Fig.\,\ref{fig:bulk_velocityfield_mu} b) do correlate visibly with the friction coefficient, i.e. correspond to the angle of repose for the given material properties.
The fluctuations of $\alpha$ in phase 3 indicate avalanches in the dilation zone along the slopes of the uplift zone. These avalanches temporarily deform the surface of the uplifted structure, which leads to momentary changes in $\alpha$. If too much material is removed by an avalanche, the material flow continues locally until the stable slope has been restored. As the uplift wedge converges onto the same shape for all $\mu$, it transitions through shapes initially associated with different $\mu$. That means, that $\alpha(\Delta h)$ for $\mu = 0.1$ can be shifted in $\Delta h$ to collapse onto the same master form as for larger $\mu$.

\subsection{Localisation of normal forces on the plate}
\begin{figure}[t]
 \centering
 \includegraphics[width=\columnwidth]{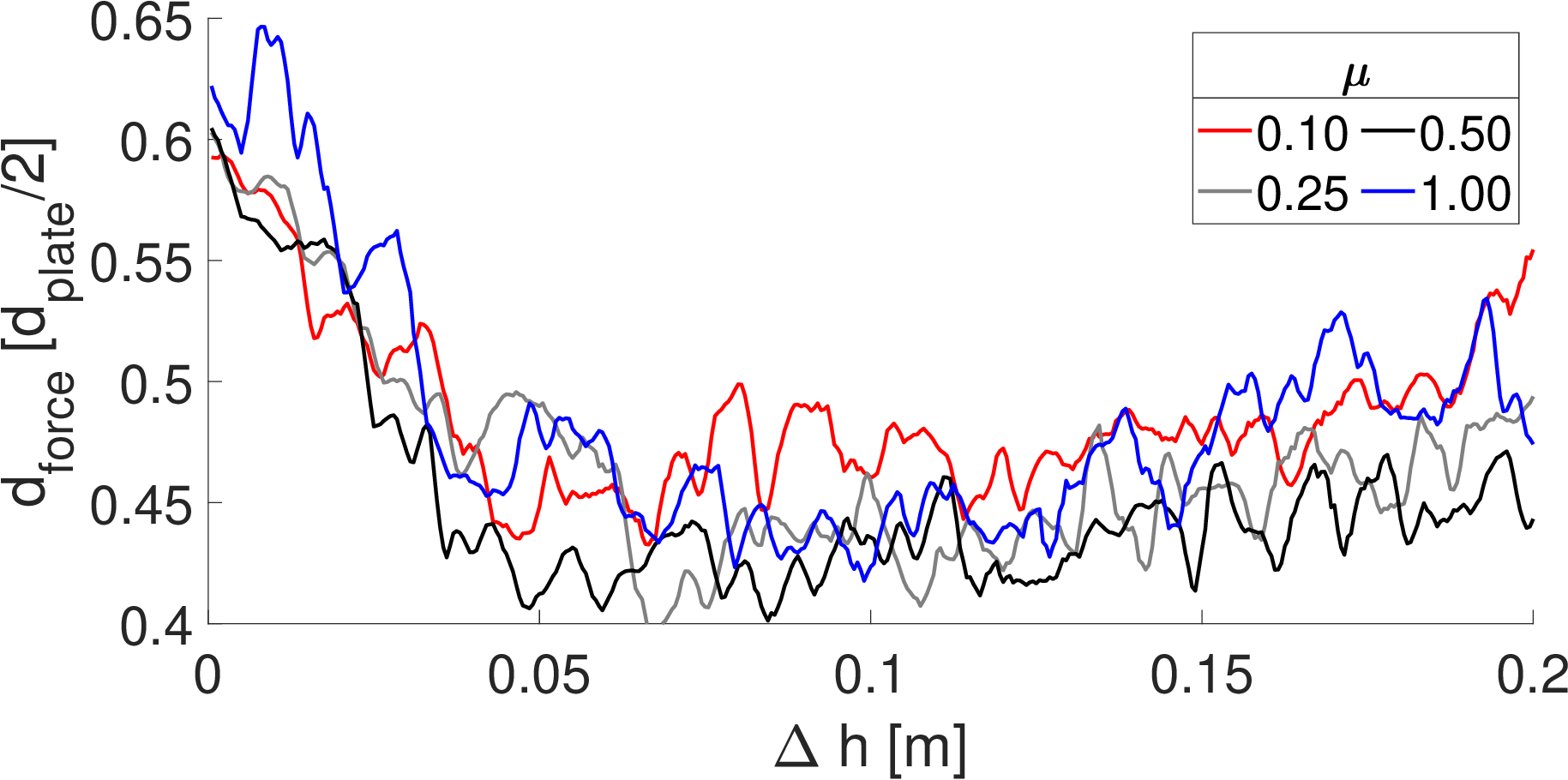}
 \caption{Distance of the force concentration on top of the plate $d_{\mathrm{force}}$ (in units of  half the plate width) with respect to the pullout distance.}
 \label{fig:anchor_force}
\end{figure}
In Fig.\,\ref{fig:bulk_fields_small} b) and section\,\ref{sec:sub_velofields} we observed that the strongest forces initially act on the edge of the plate, but move towards the plate centre after shear failure. We can quantify the distance of the bulk of the applied forces from the plate's centre by the weighted mean
\begin{equation}
 d_{\mathrm{force}} = \frac{\sum_{i}{F_{N,i}\cdot |d_i|}}{\sum_{i}{F_{N,i}}},
\end{equation}
for all normal contact forces $F_{N,i}$ acting upon the plate at distance $d_{i}$ from its centre. 

In Fig.\,\ref{fig:anchor_force}, a) we show the evolution of $d_{\mathrm{force}}$ with the pullout distance $\Delta h$. Initially, the forces act primarily on the outlying portion of the intruder, in the manner of the force chains in Fig.\,\ref{fig:bulk_fields_small} b). As the plate moves upwards, the force concentrates closer to the plate centre, i.e. is contained within the co-moving zone in Fig.\,\ref{fig:bulk_velocityfield_mu} b). 
The concurrent evolution of the force location and shape of the co-moving area also explains the evolution of the pullout resistance: As the wedge turns into a cone, the induced stresses concentrate towards a single point (the tip of the cone) in pullout direction.  This then leads to a failure of the local structure at lower total forces than for the plate or the initial wedge,  where the forces are spread outwards to the side of the aggregate. That the forces initially act on the plate edge and point outwards further enhances the pullout resistance of the aggregate through additional stabilization from the walls (compare also the higher peak pullout resistance for $\mu=1.0$ shown previously in Fig.\,\ref{fig:anchor_force_mu} and the strength of the force chains shown in Fig.\,\ref{fig:bulk_velocityfield_mu} a).
This also correlates to the findings of Vo and Nguyen\,\cite{Vo2023} that the pullout resistance is linked to the distribution of forces within the stagnant zone. We further find the evolution of the force location to be independent of the friction coefficient $\mu$, although it appears as if the forces are initially acting slightly closer to the centre of the plate if $\mu$ is small. 

\begin{figure}[t]
 \centering
 \includegraphics[width=0.7\columnwidth]{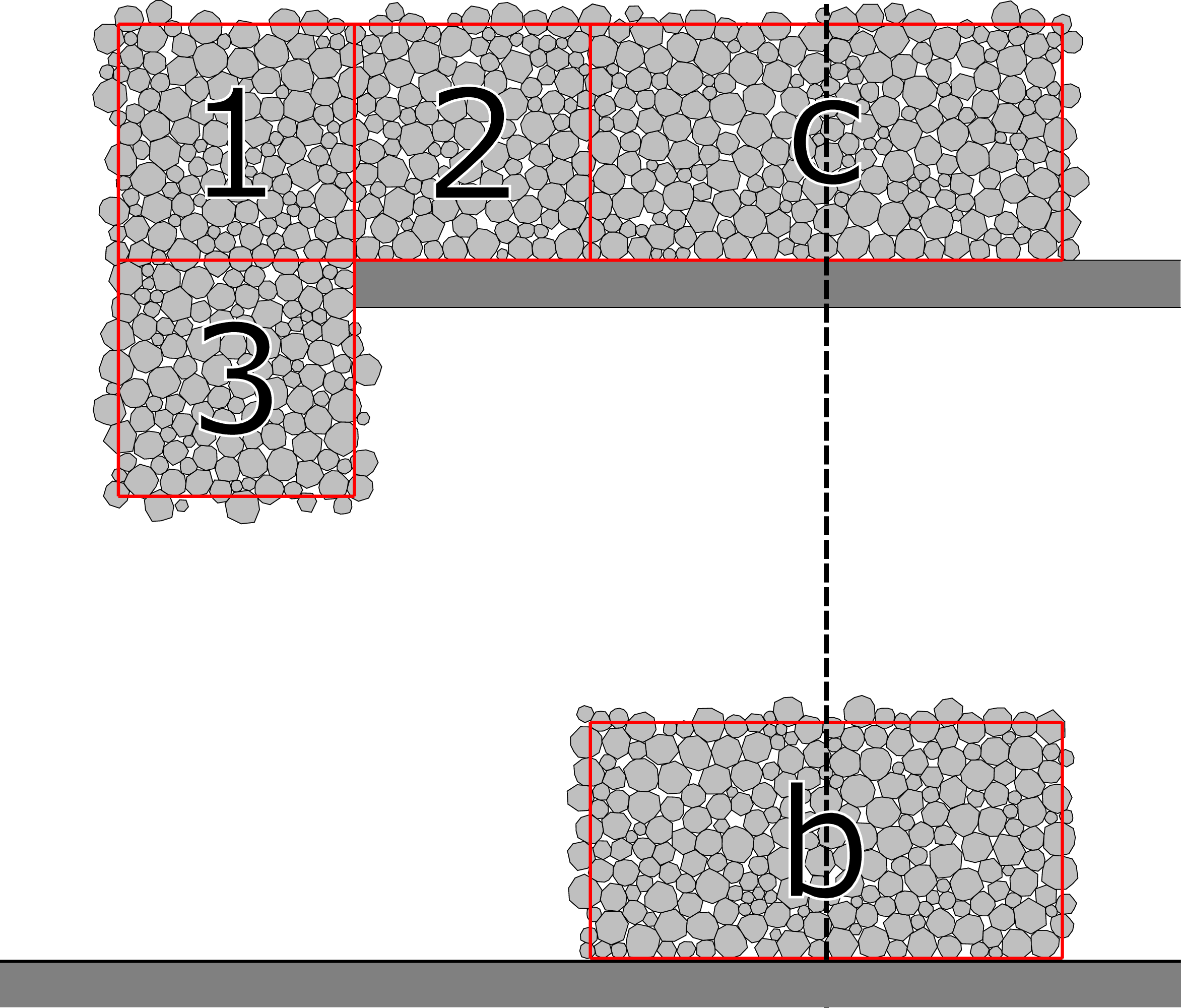}
 \caption{Location of the sampling boxes with respect to the intruder: Boxes `1', `2', `3' and `c' are co-moving with the plate. Box `b' remains fixed at the bottom of the simulation domain. The dashed line marks the centre of the intruder, respective the system.}
 \label{fig:anchor_samplebox}
\end{figure}

\subsection{Changes in the aggregate close to the plate}
As the changes in the aggregate are highly localized (see also the conclusions of Seguin et al.\,\cite{Seguin2016}) and dependent on the relative displacement of the plate, we select 3 equal square-sized sampling domains (`1'-`3') with sidelength 0.1\,[m] around the plate (inside and outside of the plate edge, above and below) that are co-moving with the plate during the pullout process. For reference we also select a co-moving sampling box (`c', 0.2$\times$0.1\,[m]) in the centre of the plate and an unmoving sampling box (`b', 0.2$\times$0.1\,[m]) at the bottom of the aggregate, see Fig.\,\ref{fig:anchor_samplebox}.

\subsubsection{Coordination number $Z$}
\begin{figure}[t]
 \centering
 \includegraphics[width=\columnwidth]{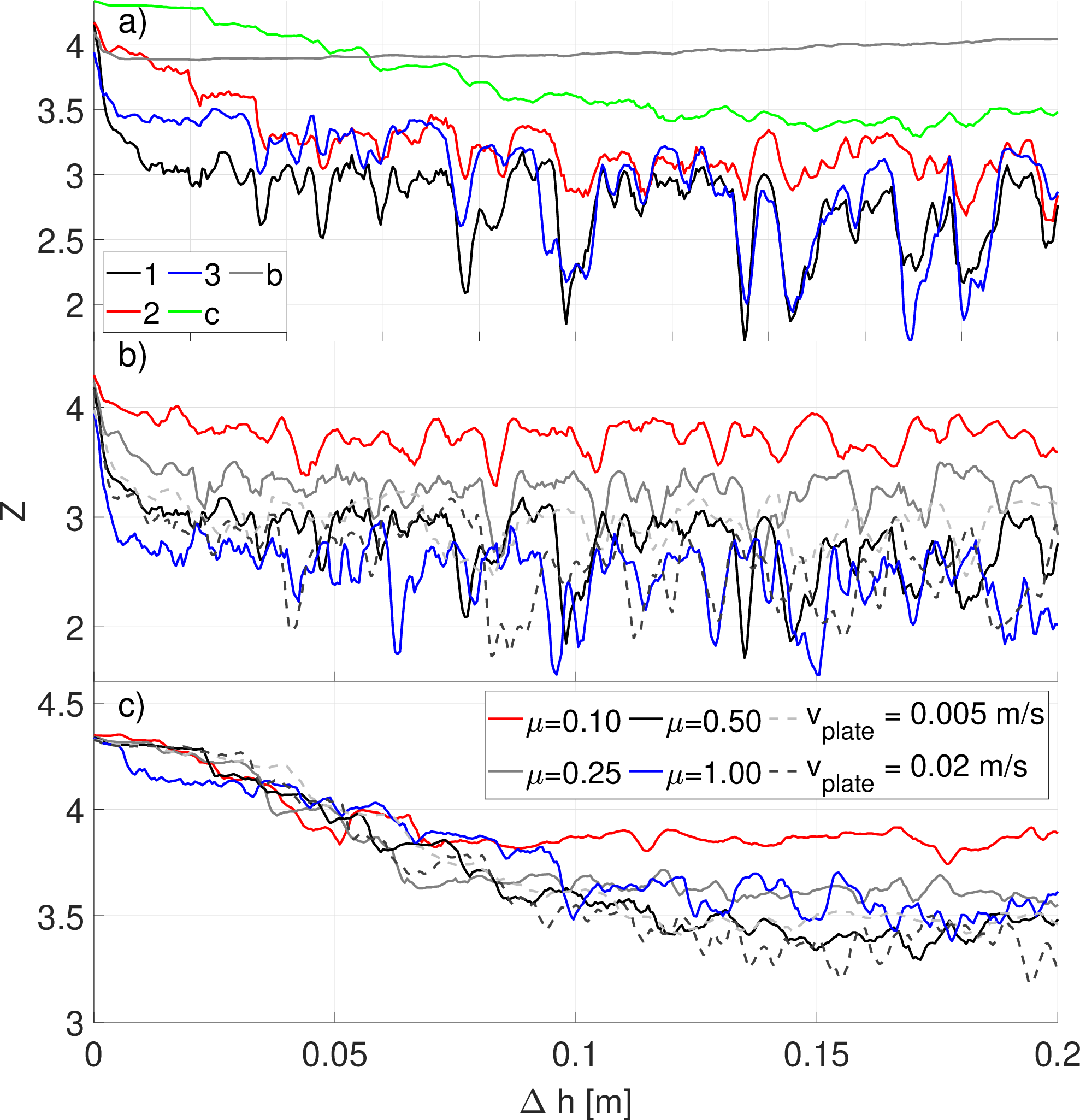}
 \caption{a) Evolution of the coordination number $Z$ with pullout distance $\Delta h$ for $\mu=0.5$ and $v_{\mathrm{plate}}=0.02$ m/s for different locations around the plate, see Fig.\,\ref{fig:anchor_samplebox}. b) Dependence of the coordination number $Z$ on the friction coefficient $\mu$ and pullout velocity $v_{\mathrm{plate}}$ in box `1' and c) box `c'.}
 \label{fig:anchor_coordnum}
\end{figure}

The porosity of the aggregate is strongly linked with its connectivity, i.e. the coordination number $Z$, defined as the mean number of contacts per particle, excluding particles with only a single contact. Figure\,\ref{fig:anchor_coordnum} a) shows the evolution of $Z$ with pullout distance $\Delta h$ for the boxes previously defined in Fig.\,\ref{fig:anchor_samplebox}. For the stationary box `b', $Z$ initially decreases as the reduced overhead load leads to local relaxation of the aggregate in the box. The coordination number then slowly increases again as new particles are deposited in the void below the plate. For the central box 'c', $Z$ decreases until it saturates at around $Z\sim 3.5$ in phase 3. At the edge of the plate (boxes `1' and `3') $Z$ drops quickly, then converges towards $Z=3$. The initial drop is the fastest for box `1' above and outside the plate and the slowest for box `2' inside the plate edge. The evolution of $Z$ is interrupted by a large decay towards $Z=2$ for the outer boxes `1' and `3'. These spikes partially correspond to spikes in the porosity $n$ for the same boxes. We understand those drops in $Z$ as large-scale avalanching events after the breakup of a temporarily jammed configuration on the plate boundaries. These events are followed by a slower increase in $Z$ as the particles outside the plate begin to jam again. In the inside box `2' $Z$ shows simultaneous, but weaker decreases compared to boxes `1' and `2', as part of the aggregate in the box remains static. 

Outside the plate the `steady state' value of $Z_{1}$ saturates according to the friction coefficient $\mu$, and appears to converge against the same value for $\mu\geq0.5$ (Fig.\,\ref{fig:anchor_coordnum} b). Our data suggests a similar behaviour in the central box $Z_{c}$ although the fluctuations for $\mu\geq0.5$ are too large to observe a clear trend for large $\mu$ (Fig.\,\ref{fig:anchor_coordnum} c).
We do not observe any influence of the pullout velocity on the coordination number, neither in the centre of the plate ($Z_{c}$) nor outside of it ($Z_{1}$).

\subsubsection{Distribution of normal contact forces and directions}\label{sec:results_forcedistrib}
\begin{figure}[t]
 \centering
 \includegraphics[width=\columnwidth]{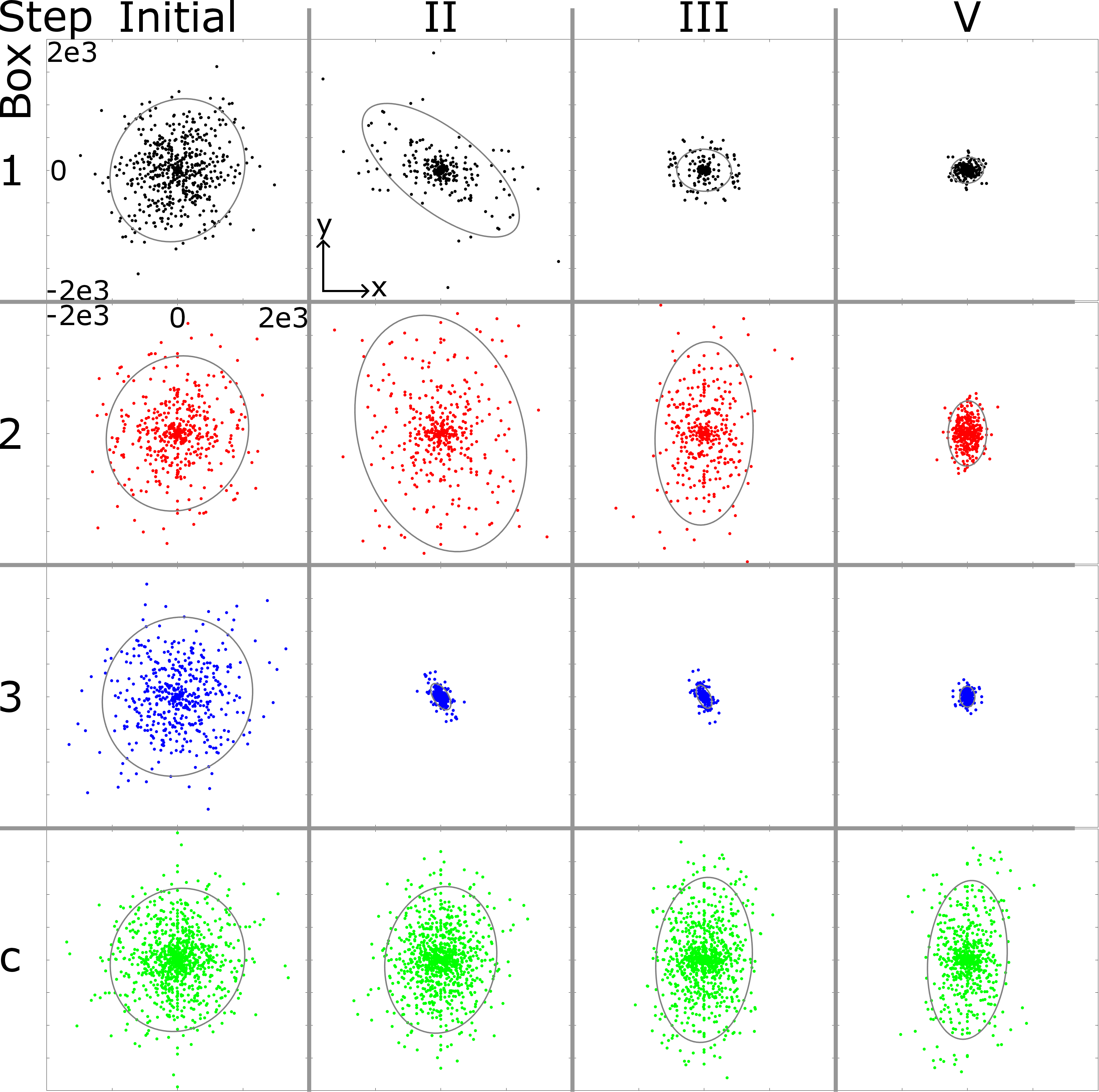}
 \caption{Distribution of contact normal force vectors within the different sampling boxes, before onset of the simulation and during different pullout steps. The elliptic curves indicate a 95\% probability contour of a 2D-Gaussian fitting on the data points. As all plots are on the same scale the fitted ellipses for box `3', steps III)-V) are barely visible within the plot.}
 \label{fig:anchor_median_FN_dir}
\end{figure}

The upward motion of the plate is a directional loading of the aggregate. In Fig.\,\ref{fig:anchor_median_FN_dir} we show the distribution of the contact normal force vectors $\bm{F}_N=F_N\bm{e}_N$. Initially $\bm{F}_N$ is isotropic and independent of the location of the sampling box, with a wide variation of strong and weak forces. As the plate begins to move, $\bm{F}_N$ evolves differently for different boxes. For all boxes on the edge of the plate, inside and outside, the overall magnitude of $F_N$ decreases. The decrease is particular significant in box `3' just below the plate, where the upwards movement of the plate and subsequent dilation of the aggregate directly above reduces its total overhead pressure.
 In step II, $\bm{F}_{N,1}$ and $\bm{F}_{N,2}$ correspond to the direction and strength of their force chains, seen earlier in Fig.\,\ref{fig:bulk_fields_small} b). As box `3' is on the right side the plate, the distribution of its normal forces is oriented from top left to bottom right. As shown previously in Figs.\,\ref{fig:bulk_fields_small} b) and\,\ref{fig:anchor_force} a) the force chains move towards the inside of the plate and start to point upwards, hence the up-down orientation of normal forces in boxes `2' and `c'. In boxes `1' and `3', the aggregate flows downwards, corresponding to the shift towards a weak uniform distribution of normal forces. Box `1' in phase 3 (e.g. step V) occasionally shows a stronger distribution in horizontal ($x$) direction, which we understand as the formation of arches, i.e. jamming. 

\begin{figure}[h!]
 \centering
 \includegraphics[width=\columnwidth]{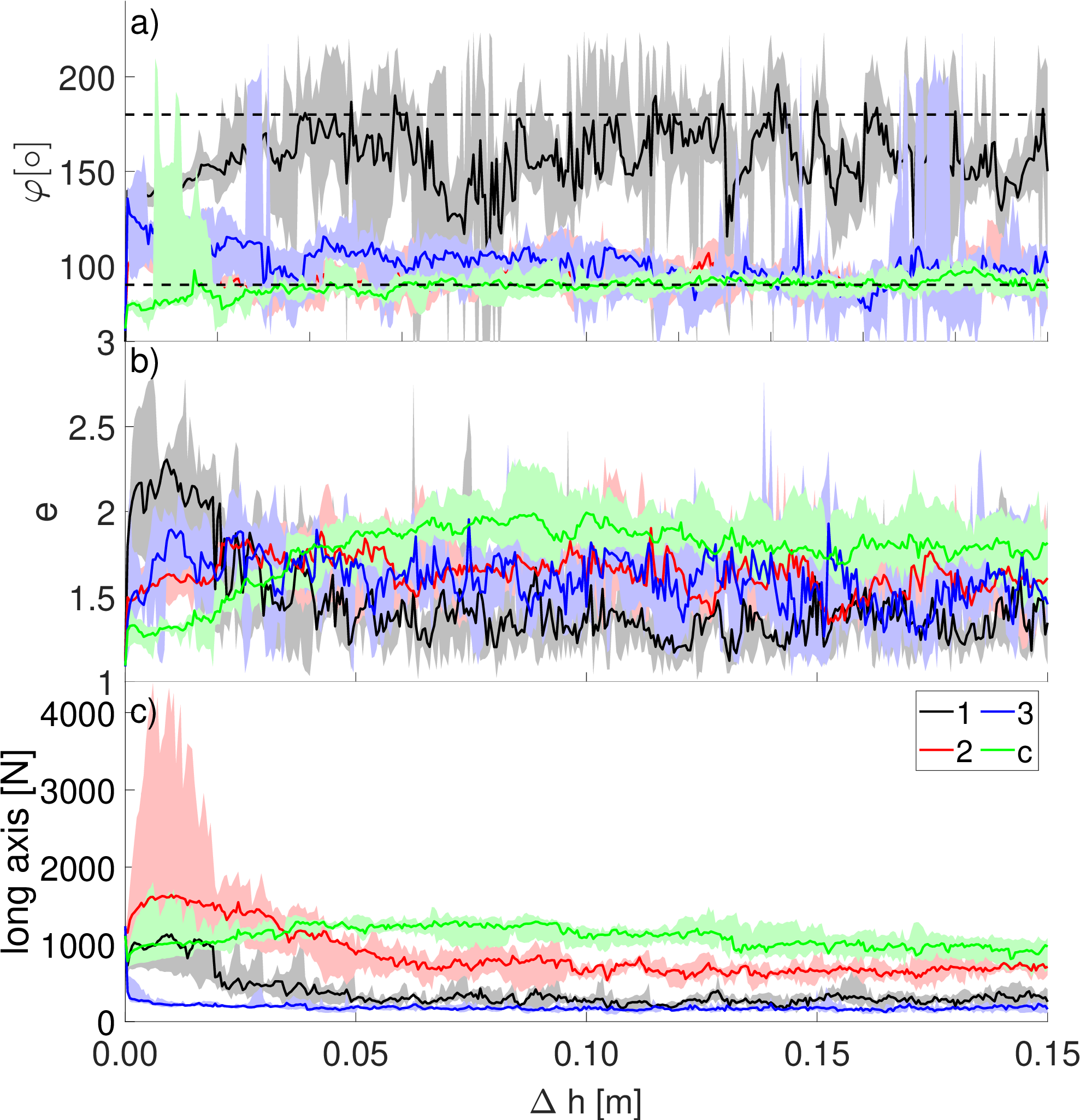}
 \caption{a) Evolution of the median orientation $\varphi$ with respect to the horizontal direction of ellipses fitted to the normal force distributions shown in Fig.\,\ref{fig:anchor_median_FN_dir}. For visual guidance we have added dashed vertical lines at $\varphi = 90^{\circ}$ and $\varphi = 180^{\circ}$. b) Evolution of the median elongation $e$ of the ellipses and c) length of the long axis. The solid lines represent the median values of $\varphi$ and $e$ across all values of $\mu$. The shadows mark the minimal and maximal values across different friction coefficients.}
 \label{fig:anchor_FN_dir_evolution}
\end{figure}

From the normal contact force data, we fit a two-dimensional Gaussian distribution with respect to the x- and y-components of $F_{\mathrm{N}}$. The 95\% probability contours of the fitted Gaussians are shown as grey ellipses in Fig.\,\ref{fig:anchor_FN_dir_evolution}. The orientation of the ellipse $\varphi$ is defined as the angle between the x-axis and its major principal axis.
The evolution of $\varphi$ with respect to the pullout distance $\Delta h$, plotted in Fig.\,\ref{fig:anchor_FN_dir_evolution} a), reinforces this observation. In boxes `2',`3', and `c', the normal forces orient themselves upwards into pullout direction (vertical direction, $\varphi=90^{\circ}$) after onset of intruder movement and then remain in that direction. In boxes `2' and `c' the orientation is due to the upward movement of the plate as source of the local loading, in box `3' the forces are oriented vertically due to gravity. On the other hand, in box `c' the force orientation varies in $150^{\circ}\leq\varphi \leq 180^{\circ}$ after shear failure. This means the force chains spread outwards horizontally from the upper plate edges with little contribution from the plate movement, but also no loading from higher particles due to gravity as the aggregate above box `1' is too dilute. An increase of $\varphi\rightarrow180^{\circ}$ is a potential indicator for onset of jamming along the force path, while conversely $\varphi\rightarrow150^{\circ}$ shows that more vertical loading occurs in box `1'. Variation of $\mu$ has little influence on the evolution of $\varphi$; it only affects when fluctuations in the curve occur.

In contrast, the elongation $e=\frac{\text{long axis}}{\text{short axis}}$ of the ellipses shows similar trends for all boxes `1',`2' and `3'at the edge of the intruder, an initial increase followed by a decrease to a steady value concurrent with the duration of phase 1. For box `c' $e$ only increases and then saturates, as interaction in the co-moving zone is primarily in loading direction only (see also the distribution of the force chains in step V shown in Fig.\,\ref{fig:bulk_fields_small} b). We wish to highlight the strong increase of $e_{1}$ before failure, which occurs in the exact same location, where the initial failure surface will develop after failure. This imbalance of forces, however, does not extend to the magnitude of forces (i.e. the length of the major axis of the fitted ellipses). Below the plate, the forces simply and quickly decrease until a steady value is reached. Above the plate, inside as outside of the plate edge, the forces initially increase due to loading induced by the plate movement. After failure the forces decay towards a steady state value for sampling boxes on the edge of the plate (`1' and `2') or saturate in the centre of the plate `c'. For most sampling boxes the magnitude of friction has little influence on the force magnitude. However, for box `2', on the inner side of the plate edge, $\mu$ controls the initial response force, with lower values of $\mu$ leading to lower forces, although the steady state forces are unaffected.

\subsubsection{Mobilization of friction}
\begin{figure}[b!]
 \centering
 \includegraphics[width=\columnwidth]{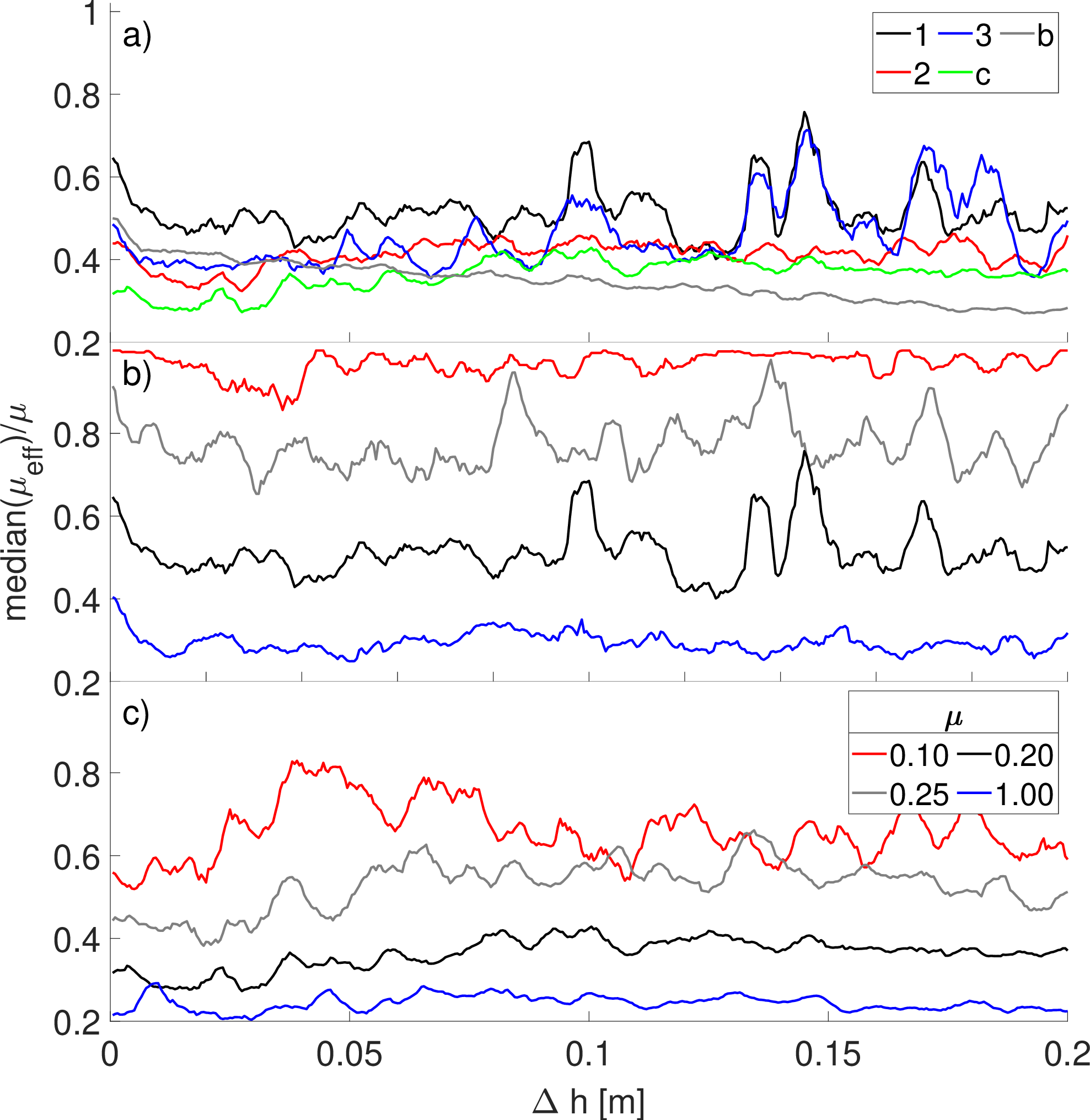}
  \caption{a) Evolution of the scaled median effective friction coefficient $\mu_{\mathrm{eff}} = \mathrm{med}(F_{T}/F_{N})/\mu$ with pullout height $\Delta h$ for $\mu=0.5$. b) Scaled effective friction coefficient $\mu_{\mathrm{eff}}$ for configurations with different values of $\mu$ in box `1' outside the plate and c) in box `c' in the centre of the plate.}
 \label{fig:anchor_median_mu}
\end{figure}
Transition between static and dynamic behaviour of granular materials is closely related to the mobilization of friction. For each contact in each sample box we compute the median effective friction coefficient $\mu_{\mathrm{eff}}= \mathrm{med}(F_{T}/F_{N})$. Figure\,\ref{fig:anchor_median_mu} a) shows the evolution of $\mu_{\mathrm{eff}}$ with the plate pullout distance $\Delta h$. 
For all boxes, except box `c' central on top of the plate, $\mu_{\mathrm{eff}}$ sharply decreases immediately after onset of the pullout as the uplift of the plate reduces the effective normal forces between the particles. 
In contrast $\mu_{\mathrm{eff},c}$ shows a minor increase in the beginning due to compaction from the uplift, followed by some decrease. In phases 2 and 3, $\mu_{\mathrm{eff},2}$ and $\mu_{\mathrm{eff},c}$ appear to be almost stable with little fluctuations. Still, as box `2' is on the edge of the plate, $\mu_{\mathrm{eff},2}$ retains larger fluctuations than box `c' as the aggregate on the edge of the plate can momentarily mobilize due to impacts and avalanching along the slopes of the uplift wedge.
Outside of the plate, $\mu_{\mathrm{eff},1}$ and $\mu_{\mathrm{eff},3}$ remain lower than the initial values, indicating that more contacts are dynamic than static. In phase 3, both $\mu_{\mathrm{eff},1}$ and $\mu_{\mathrm{eff},3}$ show significant, concurrent fluctuations, indicators of intermittent mobilization events.
For the reference box `b' at the bottom of the aggregate, $\mu_{\mathrm{eff},b}$ continuously decreases in linear manner after the initial sharp drop. The reason for this continued decay is that the aggregate was initialized without friction to generate dense initial configurations. The initial uplift of the plate then allows a slightly more relaxed configuration which remains stable at finite $\mu$ even for increasing loads $F_N$.

When varying the prescribed friction coefficient $\mu$, the effective friction coefficient $\mu_{\mathrm{eff}}$ stratifies regardless of the sampling location, see Fig.\,\ref{fig:anchor_median_mu} b)-c), where lower values of $\mu$ result in larger values $\mu_{\mathrm{eff}}/ \mu$.
For box `1' and $\mu=0.1$, friction initially decreases in phase I, but is fully mobilized afterwards, $\mu_{\mathrm{eff},1}/ \mu \sim 0.1$. 
For $\mu\geq0.5$ $\mu_{\mathrm{eff},1}$ remains below initial levels after shear softening in phase I. Additionally, we observe large fluctuations only for $\mu=0.5$, but not for $\mu = 0.1$ or $\mu = 1.0$. This may be an indicator of optimal stability in the friction-dependent competition between rolling and sliding as discussed in earlier works\,\cite{Krengel2025,Krengel2025a}.

Unlike the sampling domains at the edge of the plate, in the central box `c', friction is never fully mobilized up to the value of dynamic friction, regardless of $\mu$. Further, the effective friction $\mu_{\mathrm{eff},c}/ \mu$ initially increases and then remains larger than at the beginning. The increase is more pronounced if $\mu$ is low. We also note that fluctuations are the strongest for $\mu = 0.1$ rather than for $\mu=0.5$ as it is the case for the edge of the plate. We understand the fluctuations at low $\mu$ as the aggregate in the centre being more susceptible to repeated compression and loosening at lower decreasing shear resistance, i.e. $\mu$.

\subsubsection{Particle flow}
\begin{figure}[t]
 \centering
 \includegraphics[width=\columnwidth]{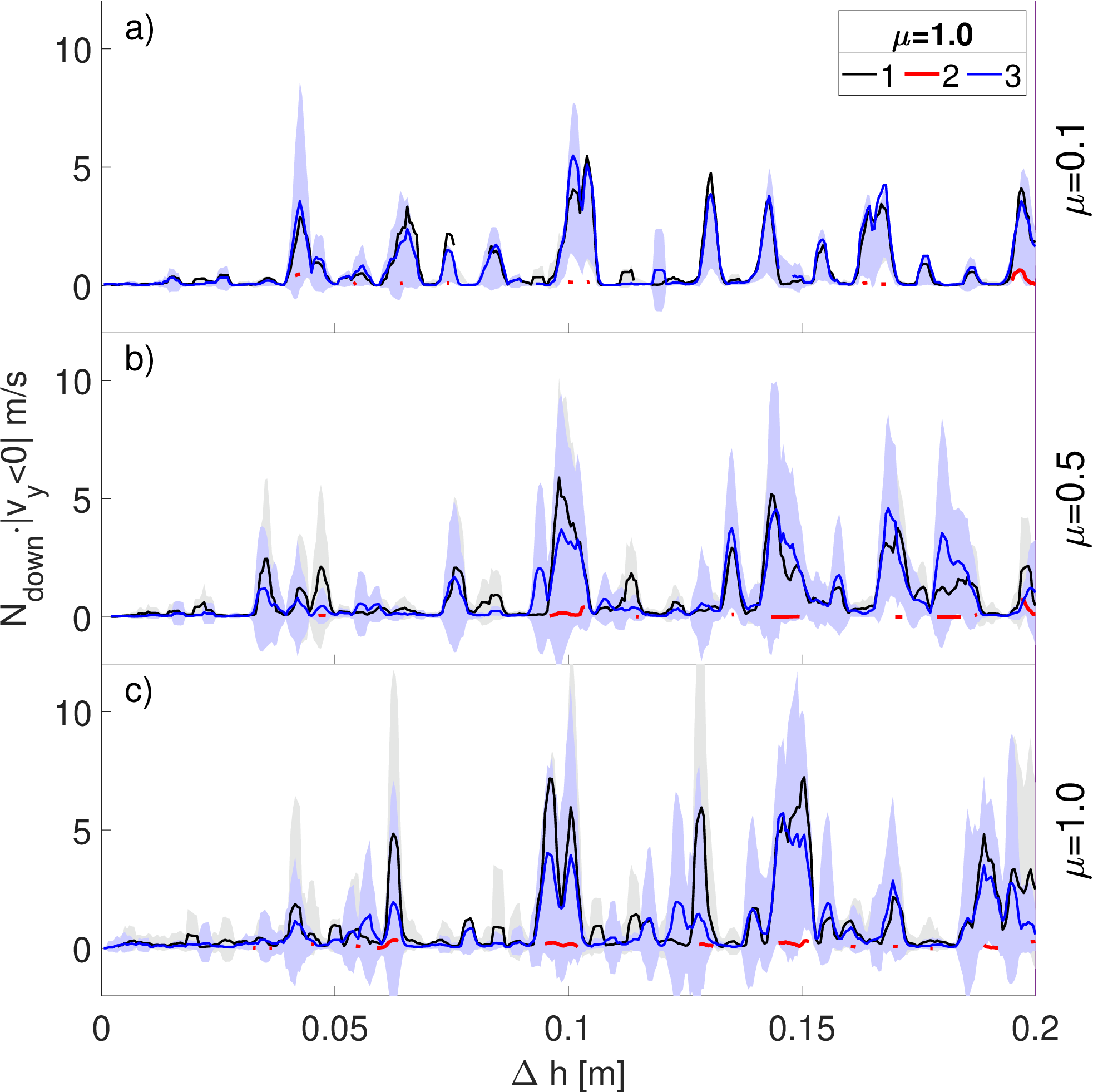}

 \caption{Mean particle flow $N_{\mathrm{down}}\cdot |v_{y}<0|$ at the edge of the plate for a) $\mu=0.1$, b) $\mu=0.5$, and c) $\mu=1.0$. The shadows mark the standard deviation around the mean value.}
 \label{fig:anchor_flowrate}
\end{figure}
As discussed previously in Sec.\,\ref{sec:results_forcedistrib}, the orientation of the normal forces is a potential indicator of jamming. We can further look at the particle flow rate, whether the aggregate is locally jammed (i.e. the mean particle flow vanishes) or flows. In Fig.\,\ref{fig:anchor_flowrate} we show the magnitude of the mean downward velocity $|v_{y}<0|$ scaled by the number of particles that are moving downwards in each box, $N_{\mathrm{down}}$. Since the aggregate in the centre of the plate is moving upwards with the plate, we omit data for box `c'. As can be seen, barely any downflow occurs in phase 1 as the aggregate has yet to dilate sufficiently enough to allow downward particle movement. After the initial shear failure, in phase 2, the material flow begins. The flow is initially weak as not enough space is available below the plate. In phase 3, once the final conical shape on top of the plate has formed , the magnitude and speed of the particle flow increases. While there is a continuous down movement of particles, the primary mass flow occurs by means of discrete events with a large variation in downward velocity, interrupted by phases of weak flows. 
We understand this pattern as a cycle of jamming between the plate boundary and the outer slope of the lower stagnant area, and subsequent avalanching once the aggregate has unjammed due to the upwards movement of the plate. In the inner box (`2') little downflow occurs compared to the outer boxes. The aggregate on the inside of the box is mostly static with respect to the plate and so any down movement occurs only during large-scale avalanches in the outer boxes (`1' and `2'), see also the red line in Fig.\,\ref{fig:anchor_flowrate}, a). In the outer boxes, avalanching occurs simultaneously but tends to decay slightly faster in the lower box, which indicates that jamming occurs on the upper plate boundary (i.e. box `1'), which cuts of the material flow in box `3', while particles can still flow into box `1'. We also note, that the mean particle flow is weaker in box `3' than in box `1',
\begin{equation}
 N_{\mathrm{down},1}\cdot v_{y,1} < N_{\mathrm{down},3}\cdot v_{y,3},
\end{equation}
as the aggregate in box `3' forms the typical deposition slopes from avalanching, and thus contains particles with little or no downward velocity (see also Fig.\,\ref{fig:bulk_fields_small}. b) and the sketch in Fig.\,\ref{fig:flow_geometry}). Interestingly, during small avalanches (in particular in phase 2) box 1 shows larger variation in the particle flow (marked by the shadows in Fig.\,\ref{fig:anchor_flowrate}), while for large avalanches, larger variation occurs in box `3' below the plate. 
The overall behaviour of the mass flow is preserved even for a variation of the friction coefficient $\mu$. For $\mu = 0.1$ (Fig.\,\ref{fig:anchor_flowrate}, b) avalanching in phase 2 is stronger, than for larger $\mu$, while the magnitude and frequency of large scale avalanches in phase 3 appear to be largely independent of $\mu$. On the other hand, the variation of the flow increases with increasing $\mu$, both for small and large scale avalanches (Fig.\,\ref{fig:anchor_flowrate}, c). We understand the independence from friction of the flow in phase 3 as yet another indicator that the final pullout behaviour is primarily controlled by the intruder geometry and less by the physical properties of the aggregate.

\subsection{Anatomy of the flow: Clogging and avalanching}
\begin{figure}[h]
 \centering
 \includegraphics[width=0.9\columnwidth]{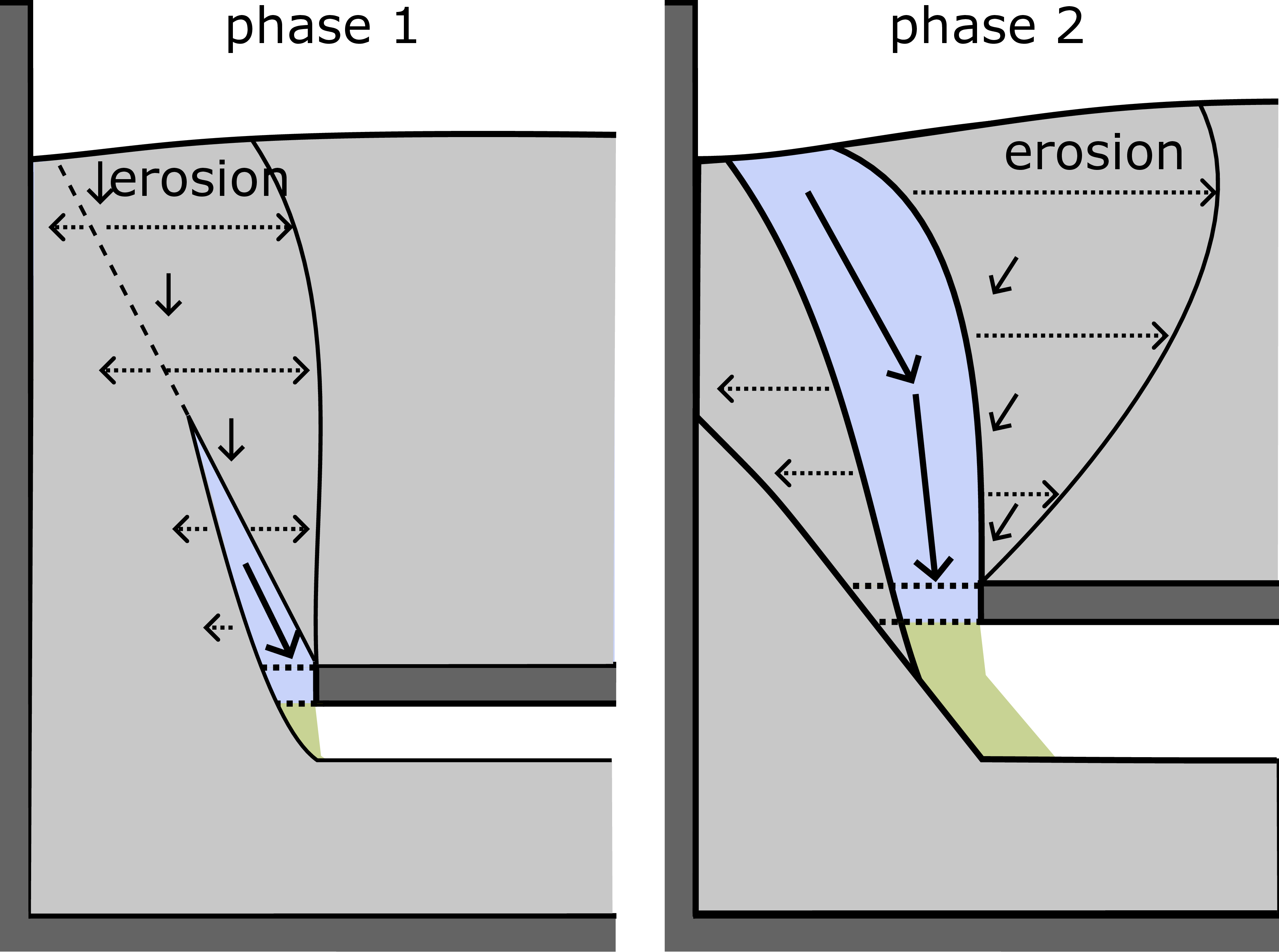}
 
 \includegraphics[width=0.9\columnwidth]{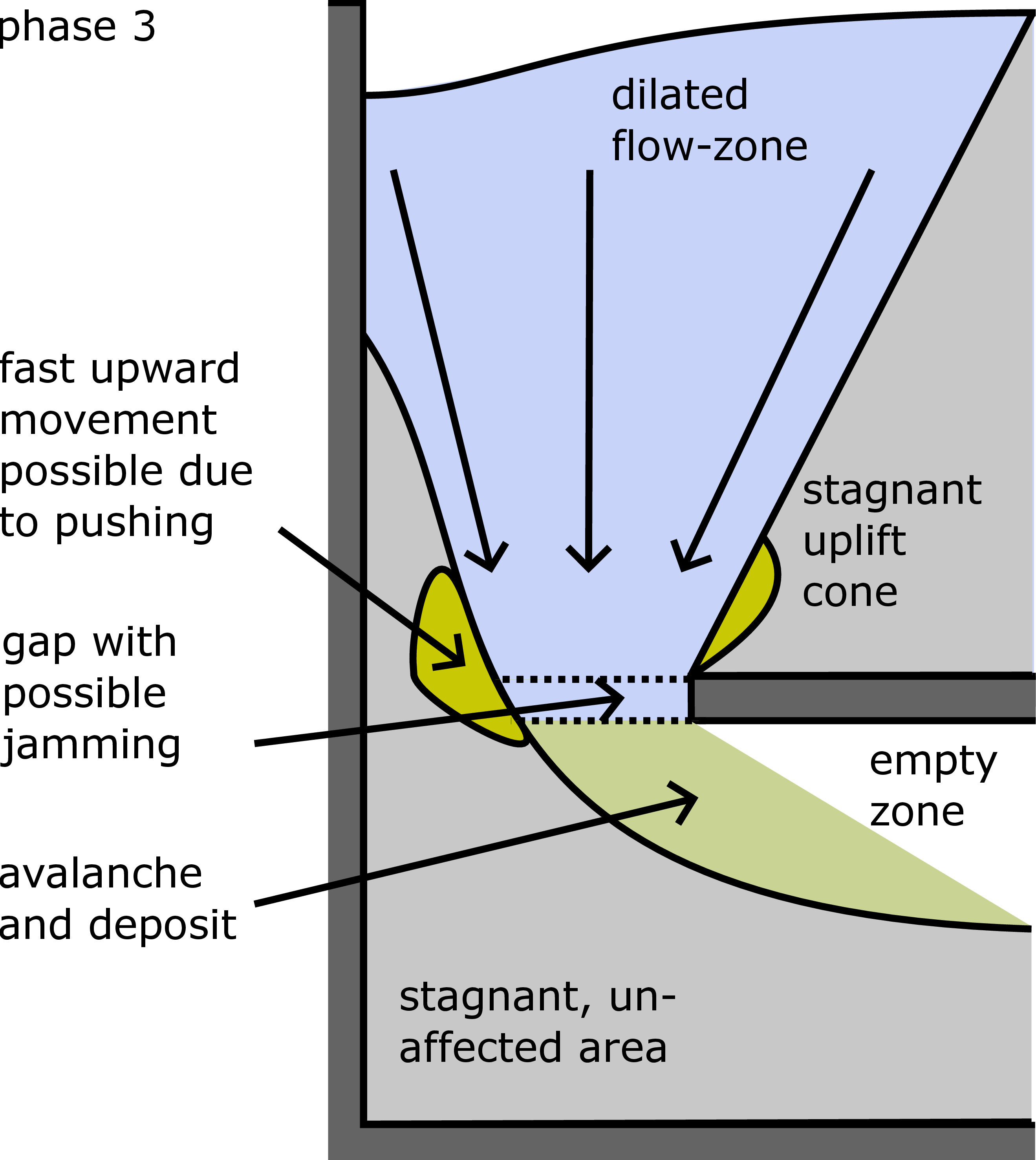}
 \caption{Sketch of the flow geometry evolution during pullout. For visualisation the height of the open space below the plate is exaggerated in phases 1 and 2.}
 \label{fig:flow_geometry}
\end{figure}

In Fig.\,\ref{fig:flow_geometry} we sketch the mechanism that drives the intruder dynamics during pullout. When the intruder begins to move upwards, it lifts the particles above via normal contacts. This zone of uplift always spreads upwards, away from the plate, 
and - depending on the relative strength of the tangential interaction forces (i.e. friction and gravity) - outwards as well. Particles below the plate, or the outside the zone of influence are unaffected by the uplift and remain in their original place, barring some small influence from relaxation due to decreasing overhead pressure. During the initial shear failure (phase 1), the aggregate dilates between the uplift zone and the outer stagnant zone. 

In this dilated zone particles begin to flow downwards under gravity. The decrease of support from underlying layers due to the dilation and the particle downflow weakens the contact stability for particles on the edge of the uplift zone (i.e. along the failure surface), leading to particles falling into the dilated zone. Thus the dilated zone spreading inwards, effectively eroding the uplifted area (phase 2). As this process proceeds from the plate upwards, this then leads to the curved failure surfaces of the uplifted area observed in previous works and in our experiment and simulation. At the same time, the particle downflow slowly erodes the outer stagnant zone as well. As the shape of the uplifted area is continuously changing during this transition phase, and the total area decreasing, the pullout resistance is decreasing non-linear in this stage.

Over time, the shape of uplifted area converges into a stable cone-like shape that does not change any further with increasing displacement (phase 3). With the effective shape of the intruder now being constant, the resistance force only depends on the pressure acting on the intruder. As the the intruder in our experiment and simulation moves upwards with a constant velocity, the pressure reduction, and therefore the decrease in resistance, is linear.
The dilated area, marked by the failure surfaces with the uplift cone and with the outer stagnant zone form a natural hopper with stable slopes and a narrow gap between the intruder and the outer stable zone. Particles eroded from the sides of the uplift material and from the outer area flow downward through this gap under the influence of gravity.
Due to the dilation, particle rolling becomes easier, i.e. particle mobility increases, which further contributes to the downflow of particles. Below the gap, the particles are deposited in typical avalanches in the empty space opened up by the plate movement. As is typical for granular avalanches, the deposit slope angle depends on the friction coefficient. As with any hopper flow, arching can induce clogging, which then interrupts the particle downflow. This jamming is then broken either by the plate moving sufficiently far upwards, or by the slopes of the `hopper' destabilizing under the acting pressure from the particles above, momentarily widening the gap and allowing particle flow again. The destabilization can also induce large scale avalanching in the dilated zone. Overall, the formation of this `hopper' means that the particle flow is not necessarily continuous, but depending on the contact stability, can also be intermittent, as expressed in the fluctuations of most quantities measured outside the edge of the plate.  

\section{Limitations}
\label{sec:limitations}
One possible concern is that the intruder is too large with respect to the size of the granular domain, and that the reaction of the aggregate to pullout is therefore affected by wall effects. Theoretical analysis and many experimental tests are conducted in a domain without finite constraints. Our data has shown no substantial influence of the boundaries for small to moderate friction coefficients ($\mu\leq0.5$). Only for large values of the friction coefficient ($\mu = 1.0$) the boundary effect becomes evident: the initial failure surface is truncated by the walls, and consequently the maximal resistance force would be affected by this truncation. However in the post failure state, the different observables of the aggregate collapse onto the same curves as for lower values of $\mu$, thus the influence of friction becomes insignificant.

A second concern is, that our simulations in this work were performed in two dimensions, which limits the degrees of freedom for the particles compared to three dimensional experiments. The overall behaviour agrees between our experiment and the simulation results, but we can expect that in three dimensions the larger surface area of the intruder will provide a greater resistance to the intruder motion. Further, the influence of the corners will likely complicate the shape and evolution of the failure surface, while the greater mobility of the particles, especially at the edge of the plate, will also affect the pullout resistance. We are planning to investigate the influence of dimensionality on the pullout in a follow-up to this study.

\section{Conclusions}
\label{sec:conclusion}
In this work, we have presented a detailed investigation into the constant-velocity pullout of a plate-like intruder from a two-dimensional granular medium by means of a polygonal discrete element simulation. We have presented our results in terms of the evolution of the bulk structure as well as the changes in the micro-mechanical properties near the intruder. Additionally, we have investigated the influence of inter-particle friction on the pullout. Below we summarize our main findings.
\begin{enumerate}
 \item  The overall behaviour of the pullout resistance can be separated into three different phases according to the local curvature: 1) Initial hardening to maximum resistance and failure, 2) a transitional stage with non-linear decrease of the resistance force, and 3) a `steady-state' equivalent with linear decrease of the resistance force. The maximum pullout resistance of the plate differs for different values of $\mu$, but collapses onto the same curve in the `steady-state'.
 \item The macroscopic fields (porosity $n$, vertical velocity $v_{y}$, cumulative rotation $\theta$ and force chains) of the aggregate show the formation of a conical structure above the plate after failure that co-moves with the plate. The formation and shape of this cone is independent of the applied friction coefficient. The slope angle of this uplift cone therefore also differs from the angles of marginal stability of a free standing granular heap.
 \item The co-moving structure above the plate evolves during pullout from an initially outward pointing wedge during shear hardening, failure and softening to a final conical conical structure in `the steady state', i.e. the effective shape of the intruder changes during the pullout. This effective shape determines the actual pullout resistance of the plate through the weight of the uplifted volume.
 \item The dilated zone between the outlying aggregate unaffected by the plate uplift and the co-moving structure forms a sort of natural hopper on the plate edge. Particles in this area move downward in discrete avalanches that are interrupted by jamming in the gap between the plate edge and the outer particles. Jamming is then broken due to the plate movement, which allows the next avalanche to form.
\end{enumerate}

The common consensus in the literature is, that the maximum pullout resistance is dependent on the weight of the uplifted volume. This volume has been shown in previous works to depend on the intruder shape\,\cite{Khatri2011,Askari2016}, size of the granular particles\,\cite{Costantino2008, Athani2017,Lehuen2020}, and increases with larger friction coefficient $\mu$,\cite{Meyerhof1968,Rowe1982,Merifield2006,Shahriar2020} as the contacts between particles on the edge of the uplift area can support more weight from above with less support from below. 
As we have shown in this work, after shear failure the shape of this uplifted volume evolves in a complex way and ultimately converges against the same conical structure regardless of the value of $\mu$, together with a collapse onto the same form for the majority of macroscopic and microscopic properties of the aggregate surrounding the intruder, including the pullout resistance. The implication of such an observation is, that the establishment of a shear zone above the intruder after material failure changes the mobility of the particles from friction dominated to geometry dominated, at the very least for quasi-static rate independent processes.

While we have only shown the convergence of the effective intruder shape towards a conical structure for originally plate-like intruders, we conjecture that different shaped intruders will show the same convergence, so as the shape of the intruder inhibits or restricts particle flow around it. Even for the commonly investigated spherical intruders, which already penetrate more easily into the particle layers ahead of its movement, formation of a small stagnant zone can be observed in experiments\,\cite{Seguin2016}, indicating the universality of this behaviour. Interestingly, for most part, the pullout of the intruder strongly resembles trapdoor tests, such as in e.g.\,\cite{Ali2020}, with only the frame of reference shifting into the intruder frame.

A possible extension of this study, which we will consider in a future work, is the application of this test to reduced gravity environments. In the presented work, the flow and downward displacement of particles around the intruder are gravity driven. This means, that the pullout of an intruder in a low or zero-gravity environment will need to push all displaced particles towards the free surface, while downward flow into the void behind the intruder will be significantly more limited. This then should lead to changes in the pullout dynamics of the intruder, possibly also in the failure modes, as, while contact forces will be weaker, they should be maintained longer, with a higher chance for geometrical jamming along the contacts.

\section*{Author contributions}
\textbf{D. Krengel:} Writing - original draft, Writing - review \& editing, Software, Methodology, Validation, Investigation, Formal analysis, Data curation, Conceptualization, Visualisation, Supervision. \\
\textbf{S. Ota:} Experiment.\\
\textbf{J. Chen:} Writing - review \& editing, Software, Data curation.\\
\textbf{S. Nomura:} Experiment, Writing - review \& editing, Supervision, Resources.\\
\textbf{H. Takahashi:} Experiment, Writing - review \& editing, Supervision.

\section*{Conflicts of interest}
There are no conflicts to declare.

\section*{Data availability}
Data is available from the corresponding author upon reasonable request.

\section*{Acknowledgements}
The authors are grateful for the support by the Grant (JPMJFR205U) from the Japan Science and Technology Agency (JST) and for the Grant-in-Aid for Scientific Research (22K04306) from the Japan Society for the Promotion of Science (JSPS). 
\begin{figure}[b]
 \centering
 \includegraphics[width=\columnwidth]{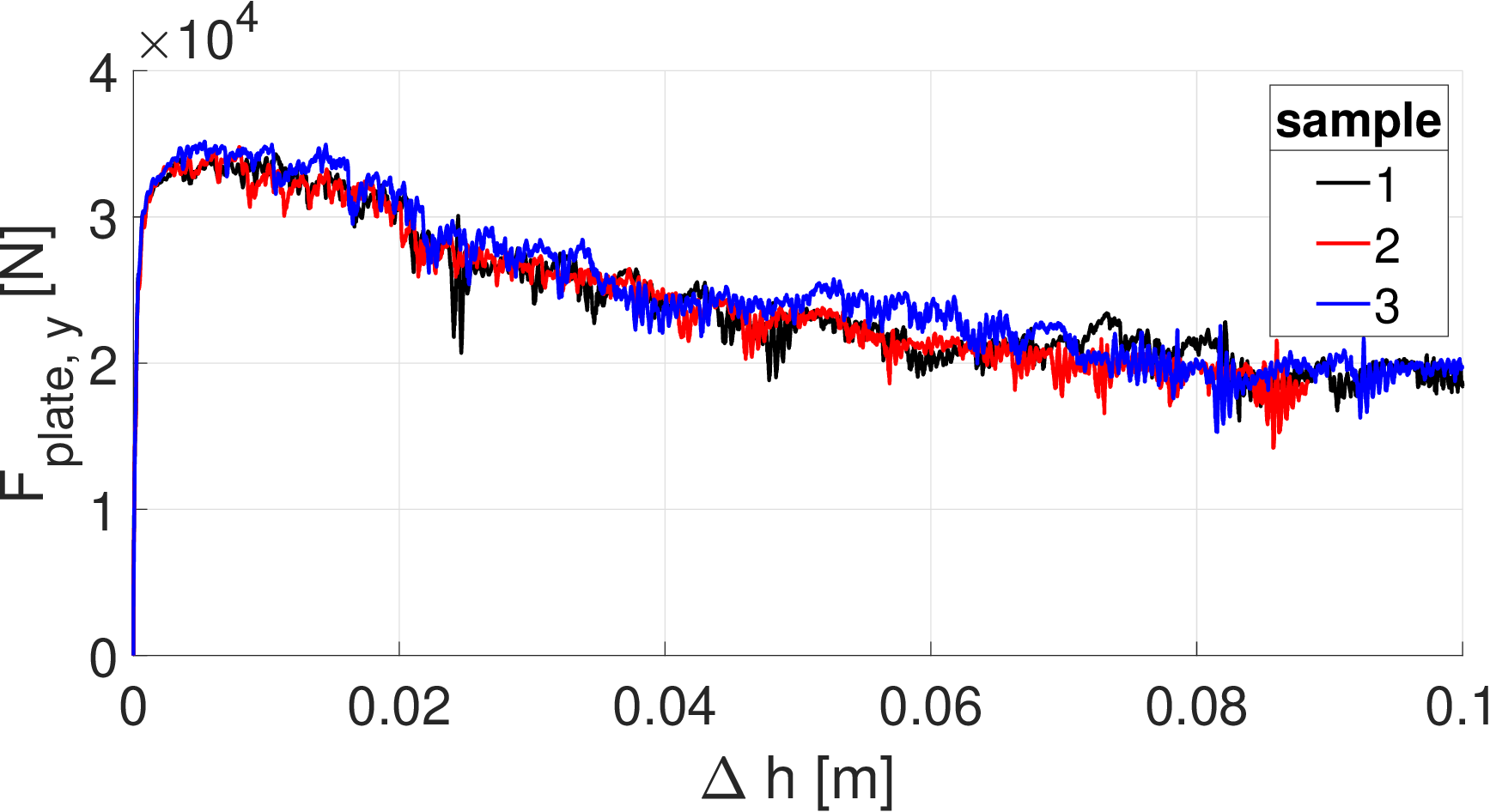}
 \caption{Resistance-displacement graph for three different initial configurations.}
 \label{fig:different_seeds}
\end{figure}

\appendix
\section*{Appendix}
\subsection*{Reproducability of the results}
\label{app:reproducability}
In order to confirm the reproducibility of our results we conducted two additional tests with different initial configurations for $\mu=0.5$ and $v_{\mathrm{plate}}=0.01$\,m/s. With the exception of the expected small variations from the different initial packing structure, the obtained macroscopic and particle-scale results were identical to the results presented in the main text, see e.g. the displacement $\Delta h$ -- pullout-resistance $F_{\mathrm{plate},y}$ diagram in Fig.\,\ref{fig:different_seeds} for the different initial configurations. We therefore assume our results to be universally valid.

\subsection*{Computing the inclination $\alpha$ of the failure surface}
\label{app:findalpha}
In this section we detail our method to determine the surface slope angle $\alpha$ of the uplifted aggregate. At first, each particle with a vertical velocity $v_{\mathrm{particle}, y}< v_{\mathrm{plate}, y}$ is removed to isolate only the particles that move upwards (Fig.\,\ref{fig:different_seeds} b). Next we remove any particle that has no force path connecting it to the intruder (Fig.\,\ref{fig:different_seeds} c). For the remaining structure we then compute the average horizontal position of the particle vertices by height to the left (red 'x'-line) and right (blue 'x'-line) of the intruder centre. This usually results in a very uneven surface in phases 2 and 3 which cannot be captured by a single parameter. To bypass this issue we instead approximate the surface by connecting the lowest horizontal mean position with the highest horizontal mean position (diamond marks) of the uplifted particles. The resulting straight line segments then enclose an angle $\alpha_{1,2}$ with the vertical direction. We define these angles so that a positive value always indicates the line segment pointing inwards and a negative value always indicates outward sloping, regardless if the slope is to the left or right of the intruder centre. The slope angle is then obtained as
\begin{equation}
 \alpha = \mathrm{mean}(\alpha_1, \alpha_2).
\end{equation}

While this approach gives a reasonable approximation of the failure surface direction in phase 1 (Fig.\,\ref{fig:different_seeds} d), it completely omits the curvature of the failure surface in phase 2 and fluctuates strongly for temporary surface variation in phase 3 (Fig.\,\ref{fig:different_seeds} c). Nevertheless, it is easily automatable and still provides a qualitative understanding how the shape of the failure surface evolves during the pullout process. Using a cumulative rotation filter would allow to isolate the failure surface much clearer than filtering by vertical velocity, but only works  for sufficiently large displacement when enough rotation has accumulated, and thus cannot be used in the earlier stages of the pullout process.

\begin{figure}[h!]
 \centering
 \includegraphics[width=0.475\columnwidth]{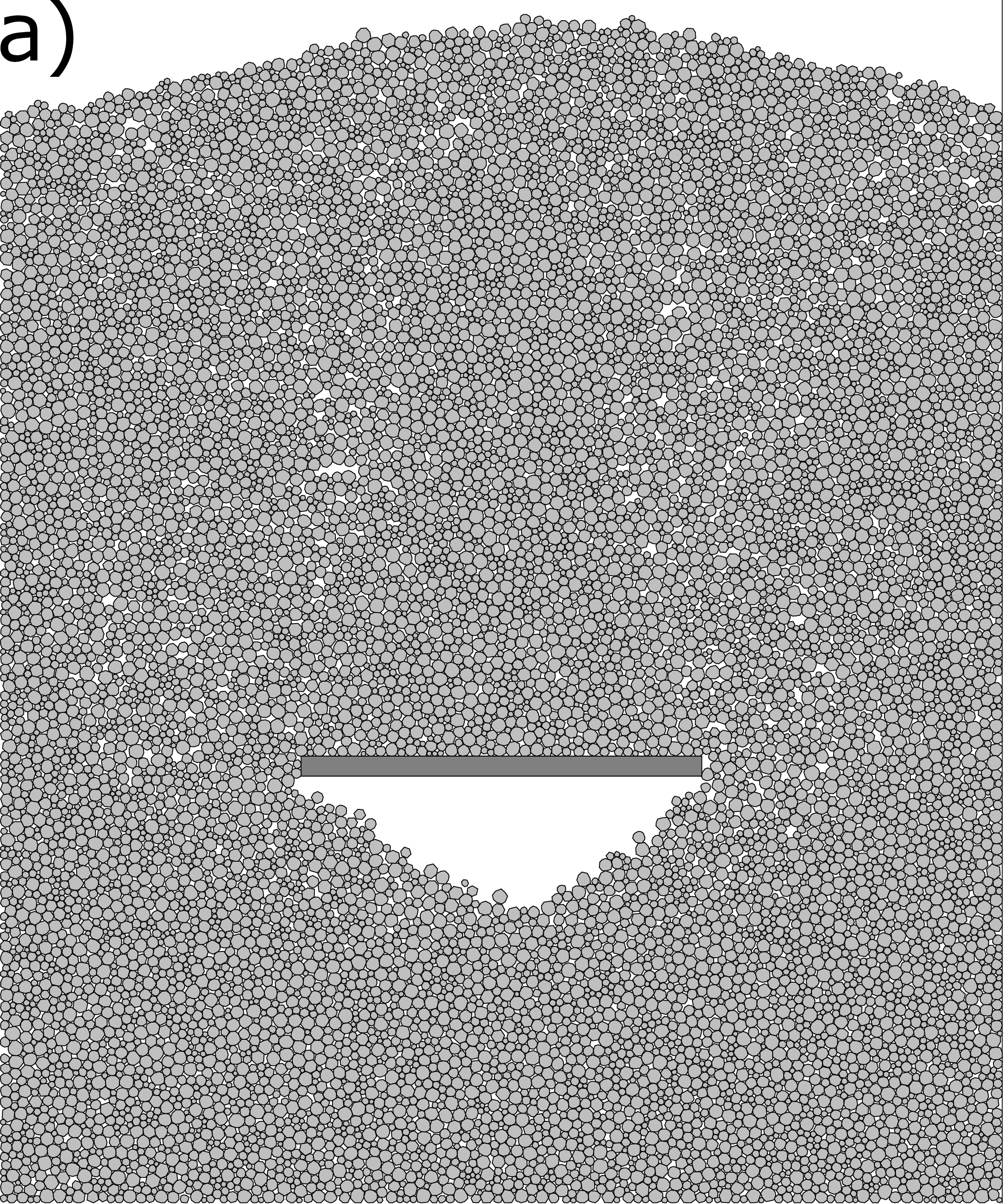}
 \hfill
 \includegraphics[width=0.475\columnwidth]{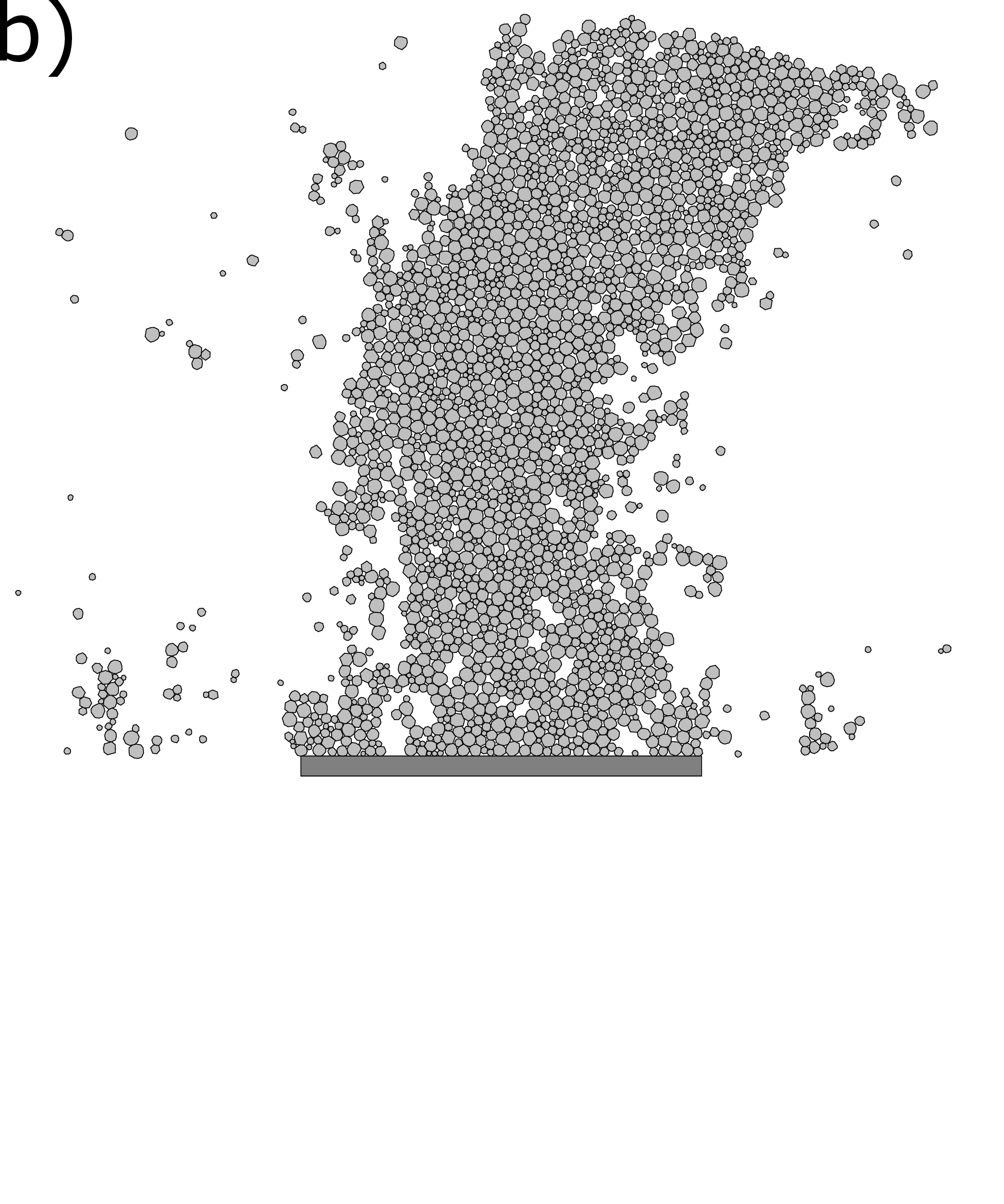}
 
 \includegraphics[width=0.475\columnwidth]{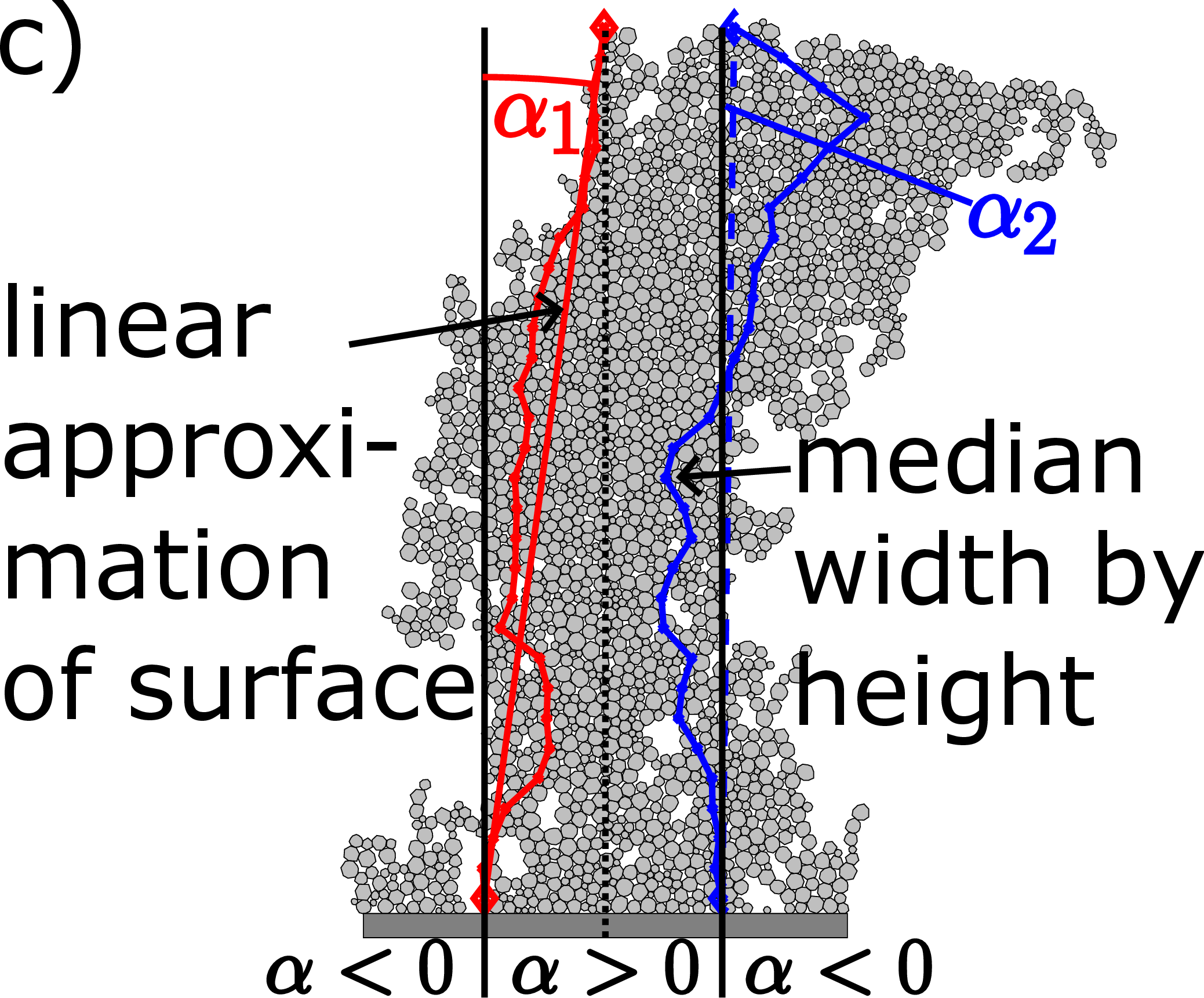}
 \hfill
 \includegraphics[width=0.475\columnwidth]{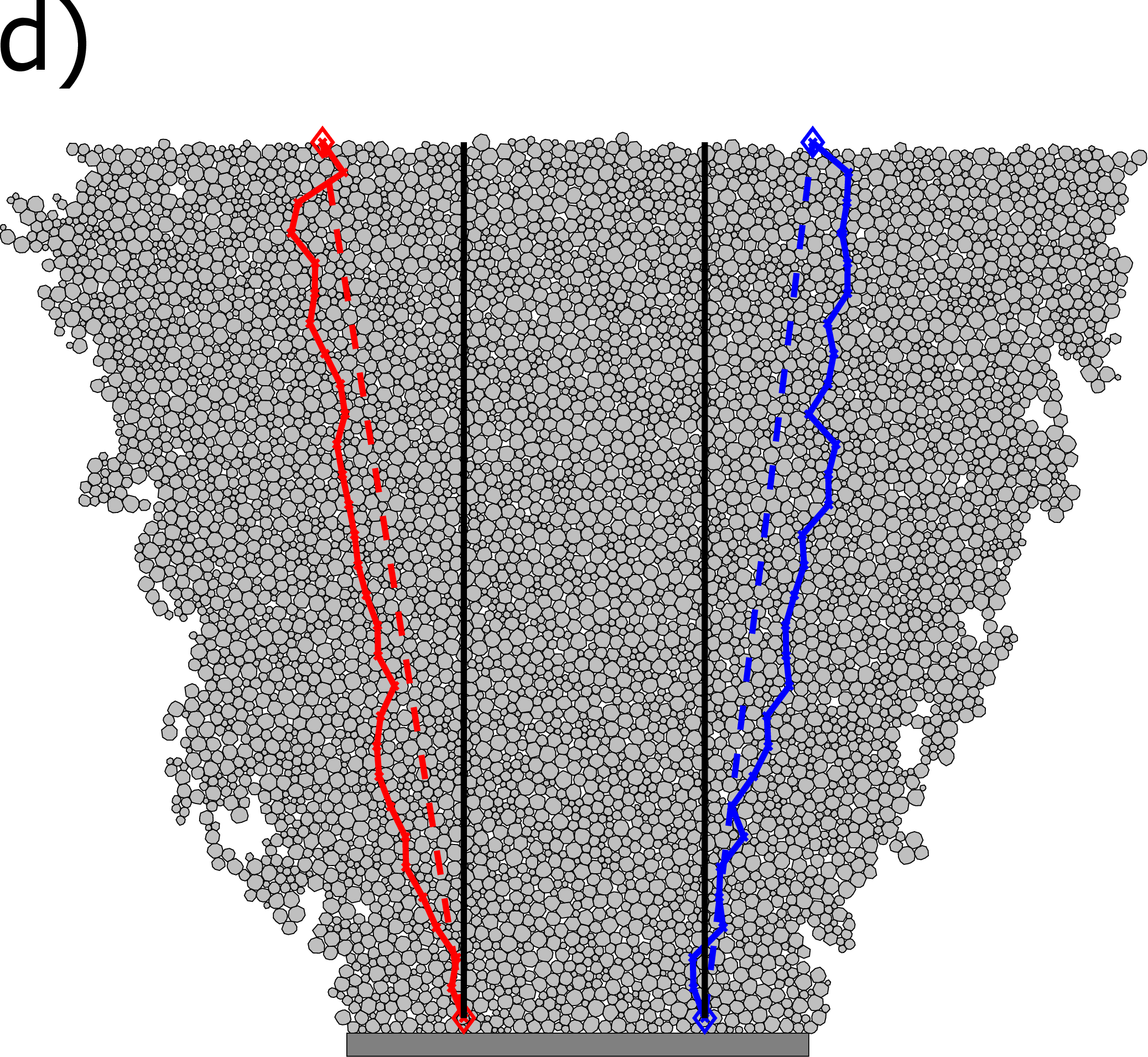}
 \caption{Filtering of the a) initial particle configuration by b) vertical velocity and c) connected force paths to the intruder. For each side of the intruder centre the slope is approximated as a straight line connecting the mean horizontal position of particles at the bottom and at the top of the remaining isolated particle structure. a)-c) In the `steady state', and d) during shear hardening.}
 \label{fig:different_seeds}
\end{figure}

\bibliography{plate_intruder} 
\end{document}